%% file: LAF.tex
\documentclass[twoside,11pt]{article}

\usepackage{jmlr2e}
\usepackage{lineno}
\usepackage[colorlinks=false]{hyperref}%<--inserisce segnalibri e riferimenti indice
\usepackage{natbib}
\usepackage{amsmath}
\usepackage{cleveref}% l
\ShortHeadings{Locally adaptive factor processes for multivariate time series}{Durante, Scarpa and Dunson}
\firstpageno{1}

\begin{document}

\title{Locally adaptive factor processes for multivariate time series}

\author{\name Daniele Durante \email durante@stat.unipd.it\\
       \addr Department of Statistical Sciences\\
       University of Padua\\
       Padua, Italy
       \AND
       \name Bruno Scarpa \email scarpa@stat.unipd.it \\
       \addr Deparment of Statistical Sciences\\
     University of Padua\\
       Padua, Italy
\AND
       \name David B. Dunson \email dunson@stat.duke.edu \\
       \addr Department of Statistical Science\\
     Duke University\\
       Durham, NC 27708-0251, USA
 }

\editor{}

\maketitle
\linenumbers

\begin{abstract}%
In modeling multivariate time series, it is important to allow time-varying smoothness in the mean and covariance process.  In particular, there may be certain time intervals exhibiting rapid changes and others in which changes are slow.  If such time-varying smoothness is not accounted for, one can obtain misleading inferences and predictions, with over-smoothing across erratic time intervals and under-smoothing across times exhibiting slow variation.  This can lead to mis-calibration of predictive intervals, which can be substantially too narrow or wide depending on the time.  We propose a locally adaptive factor process for characterizing multivariate mean-covariance changes in continuous time, allowing locally varying smoothness in both the mean and covariance matrix.  This process is constructed utilizing latent dictionary functions evolving in time through nested Gaussian processes and linearly related to the observed data with a sparse mapping.  Using a differential equation representation, we bypass usual computational bottlenecks in obtaining MCMC and online algorithms for approximate Bayesian inference.  The performance is assessed in simulations and illustrated in a financial application.
\end{abstract}


\section{Introduction}
\subsection{Motivation and setting}
\begin{figure}[t]
\centering
\includegraphics[height=9cm, width=16cm]{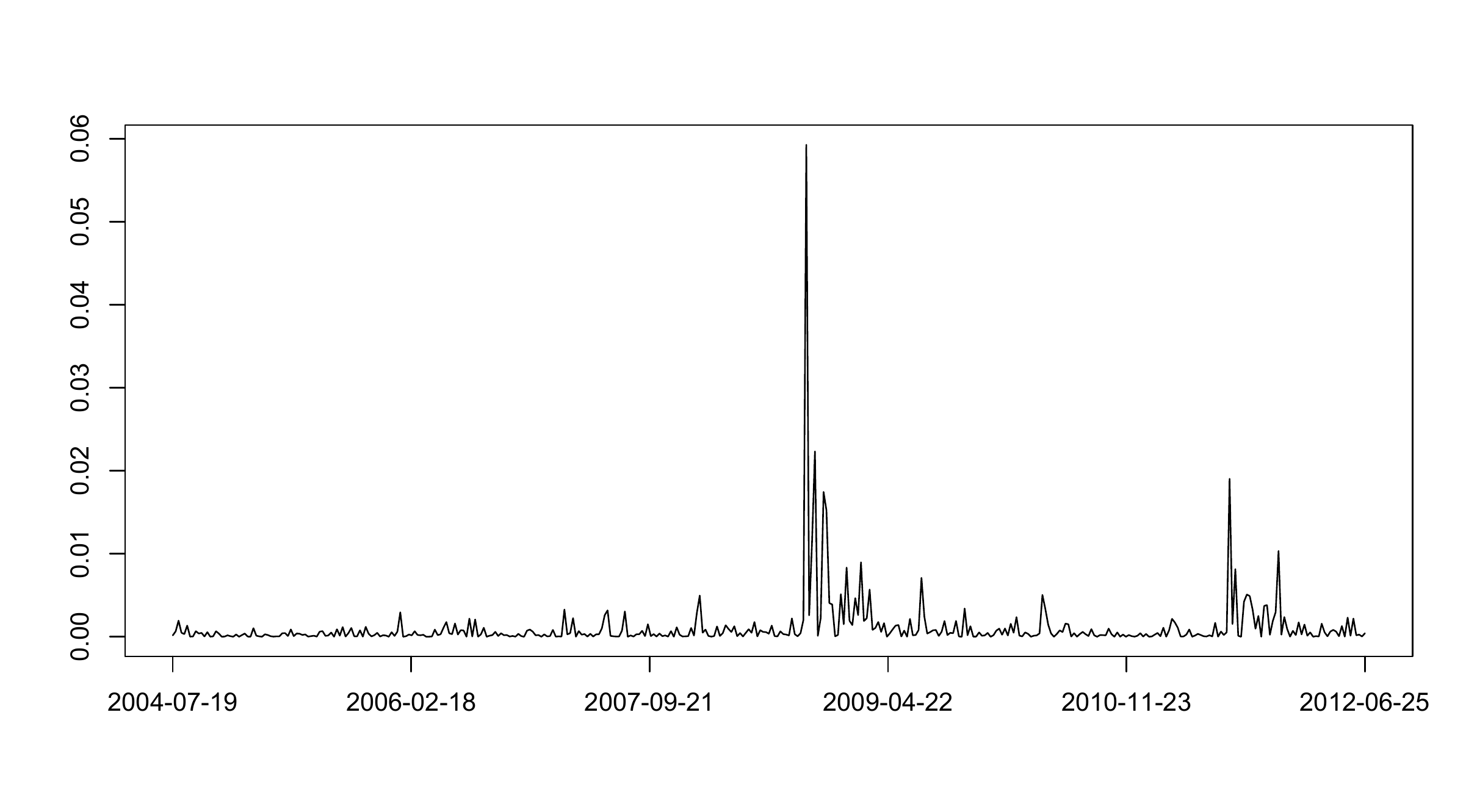}
\put (-300,225) {DAX30: Squared log returns}
\caption{\footnotesize{Squared Log-Returns of DAX30, using weekly data from $2004/07/19$, to $2012/06/25$.}}
\label{f:intro}
\end{figure}

In analyzing multivariate time series data, collected in financial applications, monitoring of influenza outbreaks and other fields, it is often of key importance to accurately characterize dynamic changes over time in not only the mean of the different elements (e.g., assets, influenza levels at different locations) but also the covariance.  As shown in Figure~\ref{f:intro}, it is typical in many domains to cycle irregularly between periods of rapid and slow change;  most statistical models are insufficiently flexible to capture such locally varying smoothness in assuming a single bandwidth parameter.  Inappropriately restricting the smoothness to be constant can have a major impact on the quality of inferences and predictions, with over-smoothing occurring during times of rapid change.  This leads to an under-estimation of uncertainty during such volatile times and an inability to accurately predict risk of extremal events. 

Let $Y_t = (Y_{t1},\ldots,Y_{tp})^T$ denote a random vector at time $t$, with $\mu(t) = \mbox{E}( Y_t )$ and $\Sigma(t) = \mbox{cov}( Y_t )$.  Our focus is on Bayesian modeling and inference for the multivariate mean-covariance stochastic process, $\Gamma = \{ \mu(t),\Sigma(t), t \in \mathcal{T}\}$ with $\mathcal{T} \subset \Re_+$.  Of particular interest is allowing locally-varying smoothness, meaning that the rate of change in the $\{ \mu(t),\Sigma(t) \}$ process is varying over time.  To our knowledge, there is no previous proposed stochastic process for a coupled mean-covariance process, which allows locally-varying smoothness.  A key to our construction is the use of latent processes, which have time-varying smoothness.   This results in a {\em locally adaptive factor} (LAF) process.
We review the relevant literature below and then describe our LAF formulation.

\subsection{Relevant literature}

There is a rich literature on modeling a $p \times 1$ time-varying mean vector $\mu(t)$, covering multivariate generalizations of autoregressive models (VAR,  e.g. \citealp{Tsay:2005}), Kalman filtering \citep{Ka:1960}, nonparametric mean regression via Gaussian processes (GP) \citep{Ra:2006}, polynomial spline \citep{Hua:2002}, smoothing spline \citep{Ha:1990} and kernel smoothing methods \citep{Wo:2011}. Such approaches perform well for slowly-changing trajectories with constant bandwidth parameters regulating implicitly or explicitly global smoothness; however, our interest is allowing smoothness to vary locally in continuous time. Possible extensions for local adaptivity include free knot splines (MARS) \citep{Fri:1991}, which perform well in simulations but the different strategies proposed to select the number and the locations of knots (stepwise knot selection \citep{Fri:1991}, Bayesian knot selection \citep{Sm:1996} or via MCMC methods \citep{Geo:1993}) prove to be computationally intractable for moderately large $p$. Other flexible approaches include wavelet shrinkage \citep{Dono:1995}, local polynomial fitting via variable bandwidth \citep{Fan:1995} and linear combination of kernels with variable bandwidths \citep{Wo:2011}. 

There is a separate literature on estimating a time-varying covariance matrix $\Sigma(t)$.  This is particular of interest in applications where volatilities and co-volatilities evolve through non constant paths.  One popular approach estimates $\Sigma(t)$ via an exponentially weighted moving average (EWMA; see, e.g., \citealp{Tsay:2005}).  This approach uses a single time-constant smoothing parameter $0<\lambda<1$, with extensions to accommodate locally-varying smoothness not straightforward due to the need to maintain positive semidefinite $\Sigma(t)$ at every time.  To allow for higher flexibility in the dynamic of the covariances, generalizations of EWMA have been proposed including the diagonal vector ARCH model (DVEC), \citep{Boll:1988} and its variant, the BEKK model \citep{Eng:1995}.  These models are computationally demanding and are not designed for moderate to large $p$.  DCC-GARCH \citep{Eng:2002} improves the computational tractability of the previous approaches through a two-step formulation. However, the univariate GARCH assumed for the conditional variances of each time series and the higher level GARCH models with the same parameters regulating the evolution of the time varying conditional correlations, restrict the evolution of the variance and covariance matrices.  PC-GARCH (\citealp{Din:1994} and \citealp{Burn:2005}) and O-GARCH \citep{Ale:2001} perform dimensionality reduction through a latent factor formulation (see also \citealp{Weide:2002}).  However, time-constant factor loadings and uncorrelated latent factors constrain the evolution of $\Sigma(t)$.  

Such models fall far short of our goal of allowing $\Sigma(t)$ to be fully flexible with the dependence between $\Sigma(t)$ and $\Sigma(t+\Delta)$ varying with not just the time-lag $\Delta$ but also with time.  In addition, these models do not handle missing data easily and tend to require long series for accurate estimation \citep{Burn:2005}.  Accommodating changes in continuous time is important in many applications, and avoids having the model be critically dependent on the time scale, with inconsistent models obtained as time units are varied.   

\citet{Wi:2010} join machine learning and econometrics efforts by proposing a model for both mean and covariance regression in multivariate time series, improving previous work of \citet{Bru:1991} on Wishart processes in terms of computational tractability and scalability, allowing a more complex structure of dependence between $\Sigma(t)$ and $\Sigma(t+\Delta)$. Specifically, they propose a continuous time Generalised Wishart Process (GWP), which defines a collection of positive semi-definite random matrices $\Sigma(t)$ with Wishart marginals. Nonparametric mean regression for $\mu(t)$ is also considered via GP priors; however, the trajectories of means and covariances inherit the smooth behavior of the underlying Gaussian processes, limiting the flexibility of the approach in times exhibiting sharp changes.  

Even for iid observations from a multivariate normal model with a single time stationary covariance matrix, there are well known problems with Wishart priors motivating a rich literature on dimensionality reduction techniques based on factor and graphical models.  There has been abundant recent interest in applying such approaches to dynamic settings.  Refer to \citet{Naka:2012} and the references cited therein for recent literature on Bayesian dynamic factor models for multivariate stochastic volatility.  Their approach allows the factor loadings to evolve dynamically over time, while including sparsity through a latent thresholding approach, leading to apparently improved performance in portfolio allocation.  They utilize a time-varying discrete-time autoregressive model, which allows the dependence in the covariance matrices 
$\Sigma(t)$ and $\Sigma(t+\Delta)$ to vary as a function of both $t$ and $\Delta$.  However, the result is an extremely richly parameterized and computationally challenging model, with selection of the number of factors proceeding by cross validation.  Our emphasis is instead on developing continuous time stochastic processes for $\Sigma(t)$ and $\mu(t)$, which accommodate locally-varying smoothness. 

\citet{Fox:2011} propose an alternative Bayesian covariance regression (BCR) model, which defines the covariance matrix as a regularized quadratic function of time-varying loadings in a latent factor model, characterizing the latter as a sparse combination of a collection of unknown Gaussian process (GP) dictionary functions.  Although their approach provides a continuous time and highly flexible model that accommodates missing data and scales to moderately large $p$, there are two limitations motivating this article.  Firstly, their proposed covariance stochastic process assumes a stationary dependence structure, and hence tends to under-smooth during periods of stability and over-smooth during periods of sharp changes.  Secondly, the well known computational problems with usual GP regression are inherited, leading to difficulties in scaling to long series and issues in mixing of MCMC algorithms for posterior computation.  

\subsection{Contribution and outline}
Our proposed LAF process instead includes dictionary functions that are generated from nested Gaussian processes (nGP) \citep{Zhu:2012}.  Such nGP reduces the GP computational burden involving matrix inversions from $O(T^3)$ to $O(T)$, with $T$ denoting the length of the time series, while also allowing flexible locally-varying smoothness.  Marginalizing out the latent factors, we obtain a stochastic process that inherits these advantages.  We also develop a different and more computationally efficient approach to computation under this new model and propose online implementation, which can accommodate streaming data.   In Section 2, we describe LAF structure with particular attention to prior specification. Section 3 explores the main features of the Gibbs sampler for posterior computation and outlines the steps for a fast online updating approach. In Section 4 we compare our model to BCR and to some of the most quoted models for multivariate stochastic volatility, through simulation studies. Finally in Section 5 an application to stock market indices across countries is examined.

\section{Locally Adaptive Factor Processes}

\subsection{ Notation and motivation} 

Our focus is on defining a novel locally adaptive factor (LAF) process for $\Gamma = \{ \mu(t),\Sigma(t), t \in \mathcal{T}\}$.  In particular, taking a Bayesian approach, we define a prior $\Gamma \sim P$, where $P$ is a probability measure over the space $\mathcal{P}$ of $p$-variate mean-covariance processes on $\mathcal{T}$.  In particular, each element of $\mathcal{P}$ corresponds to a realization of the stochastic process $\Gamma$, and the measure $P$ assigns probabilities to a $\sigma$-algebra of subsets of $\mathcal{P}$.  

Although the proposed class of LAF processes can be used much more broadly, in conducting inferences in this article, we focus on the simple case in which data consist of vectors $y_i = (y_{i1},\ldots,y_{ip})^T$ collected at times $t_i$, for $i=1,\ldots,n$.  These times can be unequally-spaced, or collected under an equally-spaced design with missing observations.  An advantage of using a continuous-time process is that it is trivial to allow unequal spacing, missing data, and even observation times across which only a subset of the elements of $y_i$ are observed.  We additionally make the simplifying assumption that 
\begin{eqnarray*}
Y_{i}\sim \mbox{N}_{p}(\mu(t_{i}),\Sigma(t_{i})).
 \label{eq:0}
\end{eqnarray*}
It is straightforward to modify the methodology to accommodate substantially different observation models.

\subsection{LAF specification} 
A common strategy in modeling of large $p$ matrices is to rely on a lower-dimensional factorization, with factor analysis providing one possible direction.  Sparse Bayesian factor models have been particularly successful in challenging cases, while having advantages over frequentist competitors in incorporating a probabilistic characterization of uncertainty in the number of factors as well as the parameters in the loadings and residual covariance.  For recent articles on Bayesian sparse factor analysis for a single large covariance matrix, refer to \citet{Bhat:2011},  \citet*{Pat:2012} and the references cited there-in.  

In our setting, we are instead interested in letting the mean vector and the covariance matrix vary flexibly over time.  Extending the usual factor analysis framework to this setting,
we say that $\Gamma = \{ \mu(t),\Sigma(t), t \in \mathcal{T}\} \sim \mbox{LAF}_{L,K}(\Theta, \Sigma_{0},\Sigma_{\xi} ,\Sigma_{A},\Sigma_{\psi}, \Sigma_{B})$ if
\vspace{-5pt}
\begin{subequations}

\begin{align}
 \mu(t)&= \Theta \xi(t) \psi(t)   \label{subeq1}\\
  \Sigma(t)&= \Theta \xi(t) \xi(t)^T \Theta^T+\Sigma_{0}  \label{subeq2}
\end{align}
\end{subequations} 
where $\Theta$ is a $p \times L$ matrix of constant coefficients, $\Sigma_{0}=\mbox{diag}(\sigma_{1}^{2},...,\sigma_{p}^{2})$, while $\xi(t)_{L \times K}$ and $\psi(t)_{K \times 1}$ are matrices comprising continuous dictionary functions evolving in time through nGP, $\xi_{lk}(t)\sim \mbox{nGP}([\Sigma_{\xi}]_{lk}=\sigma^2_{\xi_{lk}},[\Sigma_{A}]_{lk}=\sigma^2_{A_{lk}})$ and $\psi_{k }(t)\sim \mbox{nGP}([\Sigma_{\psi}]_{k}=\sigma^2_{\psi_{k}},[\Sigma_{B}]_{k}=\sigma^2_{B_{k}})$.

Restricting our attention on the generic element $\xi_{lk}(t): \mathcal{T} \rightarrow \Re$ of the matrix $\xi(t)_{L\times K}$ (the same holds for $\psi_{k}(t):\mathcal{T} \rightarrow \Re$), the nGP provides a highly flexible stochastic process on the dictionary functions whose smoothness, explicitly modeled by their $m$th order derivatives $D^{m}\xi_{lk}(t)$ via stochastic differential equations (SDEs), is expected to be centered on a local instantaneous mean function $A_{lk}(t)$, which represents a higher-level Gaussian Process (GP), that induces adaptivity to locally-varying smoothing. Specifically, we let 
\vspace{-10pt}

\begin{subequations}
\begin{align}
D^{m}\xi_{lk}(t)&=A_{lk}(t)+\sigma_{\xi_{lk}}W_{\xi_{lk}}(t), \quad m \in N, \quad m \geq 2,  \label{subeq5}\\
D^{n}A_{lk}(t)&=\sigma_{A_{lk}}W_{A_{lk}}(t), \ \quad \quad \quad \quad \ n \in N, \quad \ n \geq1,   \label{subeq6}
\end{align}
\end{subequations}
where $\sigma_{\xi_{lk}} \in \Re^+$, $\sigma_{A_{lk}} \in \Re^+$, 
$W_{\xi_{lk}}(t):\mathcal{T} \rightarrow \Re$ and $W_{A_{lk}}(t):\mathcal{T} \rightarrow \Re$ are independent Gaussian white noise processes with mean $\mbox{E}[W_{\xi_{lk}}(t)]=\mbox{E}[W_{A_{lk}}(t)]=0$, for all $t \in \mathcal{T}$, and covariance function $\mbox{E}[W_{\xi_{lk}}(t)W_{\xi_{lk}}(t')]=\mbox{E}[W_{A_{lk}}(t)W_{A_{lk}}(t')]=1$ if $t=t'$, $0$ otherwise. This formulation naturally induces a stochastic process for $\xi_{lk}(t)$ with varying smoothness, where $\mbox{E}[D^{m}\xi_{lk}(t)|A_{lk}(t)]=A_{lk}(t)$, and initialization at $t_{1}$ based on the assumption
\begin{eqnarray*}
[\xi_{lk}(t_{1}),D^{1}\xi_{lk}(t_{1}),...,D^{m-1}\xi_{lk}(t_{1})]^{T} &\sim& \mbox{N}_{m}(0,\sigma^{2}_{\mu_{lk}}I_{m})\\
\
[A_{lk}(t_{1}),D^{1}A_{lk}(t_{1}),...,D^{n-1}A_{lk}(t_{1})]^{T} &\sim& \mbox{N}_{n}(0,\sigma^{2}_{\alpha_{lk}}I_{n})\end{eqnarray*}

The Markovian property implied by SDEs in (\ref{subeq5}) and (\ref{subeq6}) represents a key advantage in terms of computational tractability as it allows  a simple state space formulation. In particular, referring to \citet{Zhu:2012} for $m=2$ and $n=1$ (this can be easily extended for higher $m$ and $n$), and for  $\delta_{i}=t_{i+1}-t_{i}$ sufficiently small, the process for $\xi_{lk}(t)$ along with its first order derivative $\xi_{lk}'(t)$ and the local instantaneous mean $A_{lk}(t)$ follow the approximated state equation
\renewcommand{\arraystretch}{0.8}
\begin{eqnarray}
\left[ \begin{array}{c}
\xi_{lk}(t_{i+1})\\
\xi'_{lk}(t_{i+1})\\
A_{lk}(t_{i+1}) \end{array} \right]=
\left[ \begin{array}{ccc}
1&\delta_{i}&0\\
0 &1&\delta_{i}\\
0 &0&1 \end{array} \right]
\left[ \begin{array}{c}
\xi_{lk}(t_{i})\\
\xi'_{lk}(t_{i})\\
A_{lk}(t_{i}) \end{array} \right]+
\left[ \begin{array}{cc}
0&0\\
1&0\\
0&1 \end{array} \right]
\left[ \begin{array}{c}
\omega_{i,\xi_{lk}}\\
\omega_{i,A_{lk}}\\
\end{array} \right],
 \label{eq:11}
\end{eqnarray}
\renewcommand{\arraystretch}{0.5}
where $[\omega_{i,\xi_{lk}}, \omega_{i,A_{lk}}]^T\sim \mbox{N}_{2}(0,V_{i,lk})$, with $V_{i,lk}=\mbox{diag}(\sigma^2_{\xi_{lk}} \delta_{i}, \sigma^2_{A_{lk}} \delta_{i})$. 

Similarly to the nGP specification for the elements in $\xi(t)$, we can represent the nested Gaussian Process for $\psi_{k}(t)$ with the following state equation
\renewcommand{\arraystretch}{0.8}
\begin{eqnarray}
\left[ \begin{array}{c}
\psi_{k}(t_{i+1})\\
\psi'_{k}(t_{i+1})\\
B_{k}(t_{i+1}) \end{array} \right]=
\left[ \begin{array}{ccc}
1&\delta_{i}&0\\
0 &1&\delta_{i}\\
0 &0&1 \end{array} \right]
\left[ \begin{array}{c}
\psi_{k}(t_{i})\\
\psi'_{k}(t_{i})\\
B_{k}(t_{i}) \end{array} \right]+
\left[ \begin{array}{cc}
0&0\\
1&0\\
0&1 \end{array} \right]
\left[ \begin{array}{c}
\omega_{i,\psi_{k}}\\
\omega_{i,B_{k}}\\
\end{array} \right]
 \label{eq:14}
\end{eqnarray}
\renewcommand{\arraystretch}{0.5}independently for $k=1,...,K$, where $[\omega_{i,\psi_{k}}, \omega_{i,B_{k}}]^T\sim \mbox{N}_{2}(0,S_{i,k})$, with $S_{i,k}=\mbox{diag}(\sigma^2_{\psi_{k}} \delta_{i}, \sigma^2_{B_{k}} \delta_{i})$. Similarly to $\xi_{lk}(t)$
\begin{eqnarray*}
[\psi_{k}(t_{1}),D^{1}\psi_{k}(t_{1}),...,D^{m-1}\psi_{k}(t_{1})]^{T} & \sim & \mbox{N}_{m}(0,\sigma^{2}_{\mu_{k}}I_{m}),\\
\
[B_{k}(t_{1}),D^{1}B_{k}(t_{1}),...,D^{n-1}B_{k}(t_{1})]^{T} & \sim & \mbox{N}_{n}(0,\sigma^{2}_{\alpha_{k}}I_{n}),
\label{eq:15}
\end{eqnarray*}
There are two crucial aspects to highlight. Firstly, this formulation allows continuous time and an irregular grid of observations over $t$ by relating the latent states at $i+1$ to those at $i$ through the distance between $t_{i+1}$ and $t_{i}$ where $i$ represents a discrete order index and $t_{i} \in \mathcal{T}$ the time value related to the $i$th observation. Secondly, compared to \citet{Zhu:2012} our approach represents an important generalization in: (i) extending the analysis to the multivariate case (i.e. $y_{i}$ is a {\em p}-dimensional vector instead of a scalar) and (ii) accommodating locally adaptive smoothing not only on the mean but also on the time-varying covariance functions.

\subsection{LAF interpretation}

Model (\ref{subeq1})-(\ref{subeq2}) can be induced by marginalizing out the $K$-dimensional latent factors vector $\eta_i$, in the model
\begin{eqnarray}
Y_{i}=\Lambda(t_{i})\eta_{i}+\epsilon_{i}, \quad \epsilon_{i}\sim \mbox{N}_{p}(0,\Sigma_{0})
 \label{eq:2}
\end{eqnarray}
where $\eta_{i}=\psi(t_{i})+\nu_{i}$ with $\nu_{i}\sim \mbox{N}_{K}(0,I_{K})$ and elements $\psi_{k }(t)\sim \mbox{nGP}(\sigma^2_{\psi_{k}},\sigma^2_{B_{k}})$ for $k=1,..., K$. In LAF formulation we assume moreover that the time-varying factor loadings matrix $\Lambda(t)$ is a sparse linear combination, with respect to the weights of the $p \times L$ matrix $\Theta$, of a much smaller set of continuous nested Gaussian Processes $\xi_{lk}(t)\sim \mbox{nGP}(\sigma^2_{\xi_{lk}},\sigma^2_{A_{lk}})$ comprising the $L\times K$, with $L<<p$, matrix $\xi(t)$. As a result
\begin{eqnarray}
\Lambda(t_{i})=\Theta\xi(t_{i})
 \label{lamb}
\end{eqnarray}
Such a decomposition plays a crucial role in further reducing the number of nGP processes to be modeled from $p \times K$ to $L \times K$ leading to a more computationally tractable formulation in which the induced $\Gamma = \{ \mu(t),\Sigma(t), t \in \mathcal{T}\}$ follows a locally adaptive factor $\mbox{LAF}_{L,K}(\Theta, \Sigma_{0},\Sigma_{\xi} ,\Sigma_{A},\Sigma_{\psi}, \Sigma_{B})$ process where
\begin{subequations}
\begin{align}
\mu(t_i)&=\mbox{E}(Y_{i} \ | \ t=t_i)=\Theta\xi(t_i)\psi(t_i) \label{subeq3}\\
\Sigma(t_i)&= \mbox{cov}(Y_{i} \ | \ t=t_i)=\Theta\xi(t_i)\xi(t_i)^T\Theta^T+\Sigma_{0}. \label{subeq4} 
\end{align}
\end{subequations}
There is a literature on using Bayesian factor analysis with time-varying loadings, but essentially all the literature assumes discrete-time dynamics on the loadings while our focus is instead on allowing the loadings, and hence the induced 
$\Gamma = \{ \mu(t),\Sigma(t), t \in \mathcal{T}\}$ processes, to evolve flexibly in continuous time.  Hence, we are most closely related to the literature on Gaussian process latent factor models for spatial and temporal data; refer, for example, to \citet{Lop:2008} and \citet{Lop:2011}.  In these models, the factor loadings matrix characterizes spatial dependence, with time varying factors accounting for dynamic changes.  

\citet{Fox:2011} instead allow the loadings matrix to vary through a continuous time stochastic process built from latent $\mbox{GP}(0,c)$ dictionary functions independently for all $l,k$, with $c$ the squared exponential correlation function having $c(x,x')=\exp(-\kappa|x-x' ||^{2}_{2})$.  In our work we follow the lead of \citet{Fox:2011} in using a nonparametric latent factor model as in (\ref{eq:2})-(\ref{lamb}), but induce fundamentally different behavior on $\Gamma = \{ \mu(t),\Sigma(t), t \in \mathcal{T}\}$ by carefully modifying the stochastic processes for the dictionary functions.

Note that the above decomposition of $\Gamma=\{ \mu(t),\Sigma(t), t \in \mathcal{T}\}$  %$\Sigma(t)$ 
is not unique.  Potentially we could constrain the loadings matrix to enforce identifiability \citep{Gew:1996}, but this approach induces an undesirable order dependence among the responses (\citealp{Ag:2000}, \citealp{West:2003}, \citealp{Lop:2004}, \citealp*{Charv:2008}). Given our focus on estimation of $\Gamma$ %$\Sigma(t)$, 
we follow \citet{Gosh:2009} in avoiding identifiability constraints, as such constraints are not necessary to ensure identifiability of the induced mean $\mu(t)$ and covariance $\Sigma(t)$.  The characterization of the class of time-varying covariance matrices $\Sigma(t)$ is proved by Lemma 2.1 of \citet{Fox:2011} which states that for $K$ and $L$ sufficiently large, any covariance regression can be decomposed as in (\ref{subeq2}). Similar results are obtained for the mean process.

\subsection{Prior Specification} 
We adopt a hierarchical prior specification approach to induce a prior $P$ on $\Gamma = \{ \mu(t),\Sigma(t), t \in \mathcal{T}\}$
with the goal of maintaining simple computation and allowing both covariances and means to evolve flexibly over continuous time. Specifically
\begin{itemize}
\item{$\Gamma |\Theta, \Sigma_{0}, \Sigma_{\xi} ,\Sigma_{A},\Sigma_{\psi}, \Sigma_{B} \sim \mbox{LAF}_{L,K}(\Theta, \Sigma_{0},\Sigma_{\xi} ,\Sigma_{A},\Sigma_{\psi}, \Sigma_{B})$  }
\item{Recalling the  nGP assumption for the elements of $\xi(t)_{L \times K}$: $\xi_{lk}(t)\sim \mbox{nGP}(\sigma^2_{\xi_{lk}},\sigma^2_{A_{lk}})$ within LAF representation, we assume for each each element $[\Sigma_{\xi}]_{lk}$ and $[\Sigma_{A}]_{lk}$ of the $L \times K$ matrices  $\Sigma_{\xi}$ and $\Sigma_{A}$ respectively, the following priors
\begin{eqnarray*}
\sigma^{2}_{\xi_{lk}} & \sim & \mbox{InvGa}(a_{\xi},b_{\xi}) \\
\sigma^{2}_{A_{lk}} & \sim & \mbox{InvGa}(a_{A},b_{A})
\label{eq:10}
\end{eqnarray*}
independently for each $(l,k)$; where $\mbox{InvGa}(a,b)$ denotes the Inverse Gamma distribution with shape $a$ and scale $b$.}
\item{Similarly, the variances $[\Sigma_{\psi}]_{k}=\sigma^{2}_{\psi_{k}}$ and $[\Sigma_{B}]_{k}=\sigma^{2}_{B_{k}}$ in the state equation representation of the nGP for each $\psi_k(t) \sim \mbox{nGP}(\sigma^2_{\psi_{k}},\sigma^2_{B_{k}})$ are assumed
\begin{eqnarray*}
\sigma^{2}_{\psi_{k}} & \sim & \mbox{InvGa}(a_{\psi},b_{\psi}) \\
\sigma^{2}_{B_{k}} & \sim & \mbox{InvGa}(a_{B},b_{B})
\label{eq:16}
\end{eqnarray*} 
independently for each $k$.}
\item{
To address the issue related to the selection of the number of dictionary elements a shrinkage prior is proposed for $\Theta$. In particular, following \citet{Bhat:2011} we assume:  
\begin{eqnarray}
\theta_{jl}|\phi_{jl},\tau_{l}\sim \mbox{N}(0,\phi_{jl}^{-1}\tau_{l}^{-1}) \quad \phi_{jl}\sim \mbox{Ga}(3/2,3/2) \nonumber\\
\vartheta_{1}\sim \mbox{Ga}(a_{1},1), \quad \vartheta_{h}\sim\mbox{ Ga}(a_{2},1), h \geq2, \quad \tau_{l}=\prod_{h=1}^{l}\vartheta_{h}
\label{eq:12}
\end{eqnarray}
Note that if $a_{2}>1$ the expected value for $\vartheta_{h}$ is greater than $1$. As a result, as $l$ goes to infinity, $\tau_{l}$ tends to infinity shrinking $\theta_{jl}$ towards zero. This leads to a flexible prior for $\theta_{jl}$ with a local shrinkage parameter $\phi_{jl}$ and a global column-wise shrinkage factor $\tau_{l}$ which allows many elements of $\Theta$ being close to zero as $L$ increases.}
\item{Finally for the variances of the error terms in vector $\epsilon_{i}$, we assume the usual inverse gamma prior distribution. Specifically 
\begin{eqnarray*}
\sigma_{j}^{-2} \sim \mbox{Ga}(a_{\sigma},b_{\sigma})
\label{eq:13}
\end{eqnarray*}
independently for each $j=1,...,p$.
}
\end{itemize}

\section{ Posterior Computation}
For a fixed truncation level $L^{*}$ and a latent factor dimension $K^{*}$, the algorithm for posterior computation alternates between a simple and efficient simulation smoother step \citep{Durb:2002} to update the state space formulation of the nGP in LAF prior, and standard Gibbs sampling steps for updating the parametric component parameters from their full conditional distributions. 
\subsection{Gibbs Sampling}
We outline here the main features of the algorithm for posterior computation based on observations $(y_{i},t_{i})$ for $i=1,...,T$, while the complete algorithm is provided in the Appendix.
\begin{description}
\item {A.} Given $\Theta$ and $\{\eta_{i}\}_{i=1}^{T}$, a multivariate version of the MCMC algorithm proposed by \citet{Zhu:2012} draws posterior samples from each dictionary element's function $\{\xi_{lk}(t_{i})\}_{i=1}^{T}$, its first order derivative $\{\xi'_{lk}(t_{i})\}_{i=1}^{T}$, the corresponding instantaneous mean $\{A_{lk}(t_{i})\}_{i=1}^{T}$, the variances in the state equations $\sigma_{\xi_{lk}}^{2}$, $\sigma_{A_{lk}}^{2}$ and the variances of the error terms in the observation equation $\sigma_{j}^{2}$ with $j=1,...,p$.
\item {B.} Given $\Theta$, $\{\sigma_{j}^{-2}\}_{j=1}^{p}$, $\{ y_{i} \}_{i=1}^{T}$ and $\{ \xi(t_{i})\}_{i=1}^{T}$  we implement a block sampling of $\{\psi(t_{i})\}_{i=1}^{T}$, $\{\psi'_{k}(t_{i})\}_{i=1}^{T}$, $\{B_{k}(t_{i})\}_{i=1}^{T}$ ,$\sigma_{\psi_{k}}^{2}$, $\sigma_{B_{k}}^{2}$ and $\nu_i$ following a similar approach as in step A.
\item {C.} Conditioned on $\{y_{i}\}_{i=1}^{T}$, $\{\eta_{i}\}_{i=1}^{T}$, $\{\sigma_{j}^{-2}\}_{j=1}^{p}$ and $\{\xi(t_{i})\}_{i=1}^{T}$, and recalling the shrinkage prior for the elements of $\Theta$ in (\ref{eq:12}), we update $\Theta$, each local shrinkage hyperparameter $\phi_{jl}$ and the global shrinkage hyperparameters $\tau_{l}$ following the standard conjugate analysis.
\item {D. Given the posterior samples from $\Theta$, $\Sigma_{0}$, $\{ \xi(t_{i})\}_{i=1}^{T}$ and $\{\psi(t_{i})\}_{i=1}^{T}$ the realization of LAF process for $\{\mu(t_i),\Sigma(t_i), t_i \in \mathcal{T} \}$ conditioned on the data $\{y_{i}\}_{i=1}^{T}$ is 
\begin{eqnarray*}
\mu(t_i)&=&\Theta\xi(t_i)\psi(t_i)\\
\Sigma(t_i)&= &\Theta\xi(t_i)\xi(t_i)^T\Theta^T+\Sigma_{0}.
\end{eqnarray*}}
\end{description}

\subsection{Hyperparameter interpretation} 
We now focus our attention on the hyperparameters of the priors for $\sigma^{2}_{\xi_{lk}}$, $\sigma^{2}_{A_{lk}}$, $\sigma^{2}_{\psi_{k}}$ and $\sigma^{2}_{B_{k}}$. Several simulation studies have shown that the higher the variances in the latent state equations, the better our formulation accommodates locally adaptive smoothing for sudden changes in $\Gamma$. A theoretical support for this data-driven consideration can be identified in the connection between the nGP and the nested smoothing splines. It has been shown by \citet{Zhu:2012} that the posterior mean of the trajectory $U$ with reference to the problem of nonparametric mean regression under the nGP prior can be related to the minimizer of the equation
\begin{eqnarray*}
\frac{1}{T}\sum_{i=1}^{T}(y_{i}-U(t_{i}))^{2}+\lambda_{U}\int_{\mathcal{T}}(D^{m}U(t)-C(t))^2 dt+\lambda_{C}\int_{\mathcal{T}}(D^{n}C(t))^2 dt,
\label{eq:17}
\end{eqnarray*}
where $C$ is the locally instantaneous function and $\lambda_{U} \in \Re^{+}$ and $\lambda_{C} \in \Re^{+}$ regulate the smoothness of the unknown functions $U$ and $C$ respectively, leading to less smoothed patterns when fixed at low values. The resulting inverse relationship between these smoothing parameters and the variances in the state equation, together with the results in the simulation studies, suggest to fix the hyperparameters in the Inverse Gamma prior for $\sigma^{2}_{\xi_{lk}}$, $\sigma^{2}_{A_{lk}}$, $\sigma^{2}_{\psi_{k}}$ and $\sigma^{2}_{B_{k}}$ so as to allow high variances in the case in which the time series analyzed are expected to have strong changes in their covariance (or mean) dynamic. A further confirmation of the previous discussion is provided by the structure of the simulation smoother required to update the dictionary functions in our Gibbs Sampling for posterior computation. More specifically, the larger the variances of $\{\omega_{i,\xi_{lk}}\}_{i=1}^{T}$,  $\{\omega_{i,A_{lk}}\}_{i=1}^{T}$ and  $\{\omega_{i,\psi_{k}}\}_{i=1}^{T}$,  $\{\omega_{i,B_{k}}\}_{i=1}^{T}$ in the state equations, with respect to those of the vector of observations $\{y_{i}\}_{i=1}^{T}$, the higher is the weight associated to innovations in the filtering and smoothing techniques, allowing for less smoothed patterns both in the covariance and mean structures (see \citealp{Durb:2002}).

In practical applications, it may be useful to obtain a first estimate of 
$\tilde\Gamma=\{\tilde{\mu}(t), \tilde{\Sigma}(t)\}$
%the covariance matrix $\tilde{\Sigma}(t)$ and the mean vector $\tilde{\mu}(t)$  
to set the hyperparameters. More specifically, $\tilde{\mu}_{j}(t_{i})$ can be the output of a standard moving average on each time series $y_{j}=[y_{j1},...,y_{jT}]$, while $\tilde{\Sigma}(t_{i})$ can be obtained by a simple estimator, such as the EWMA procedure. With these choices, the recursive equation
\begin{eqnarray*}
\tilde{\Sigma}(t_{i})=(1-\lambda)\{[y_{i-1}-\tilde{\mu}(t_{i-1})][y_{i-1}-\tilde{\mu}(t_{i-1})]^T\}+\lambda \tilde{\Sigma}(t_{i-1})
\label{eq:18}
\end{eqnarray*}
become easy to implement.

%Another aspect to point out is that the nGP prior represents a generalization of the GP prior for polynomial smoothing spline, including the latter as a special case when the variances of the local instantaneous means in the state space equations, together with those of the initial distributions for the latent states at $t_{1}$ approaches to $0$. Such a result plays a crucial role to ensure the flexibility of our approach in modeling both varying smoothing processes, both smooth processes when required.

\subsection{Online Updating}
The problem of online updating represents a key point in multivariate time series with high frequency data. Referring to our formulation, we are interested in updating an approximated posterior distribution for $\Gamma_{T+H}= \{\mu(t_{T+h}), \Sigma(t_{T+h}), h=1, ..., H\}$
%$\Sigma(t_{T+h})$ and $\mu(t_{T+h})$ with $h=1, ..., H$ 
once a new vector of observations $\{y_{i}\}_{i=T+1}^{T+H}$ is available, instead of rerunning posterior computation for the whole time series. 

Using the posterior estimates of the Gibbs sampler based on observations available up to time $T$, $\{y_{i}\}_{i=1}^{T}$, it is easy to implement (see in Appendix) a highly computationally tractable online updating algorithm which alternates between steps A, B and D outlined in the previous section for the new set of observations, and that can be initialized at $T+1$ using the one step ahead predictive distribution for the latent state vectors in the state space formulation.

Note that the initialization procedure for latent state vectors in the algorithm depends on the sample moments of the posterior distribution for the latent states at $T$. As is known for Kalman smoothers (see, e.g., \citealp{Durb:2001}), this could lead to computational problems in the online updating due to the larger conditional variances of the latent states at the end of the sample (i.e., at $T$). To overcome this problem, we replace the previous assumptions for the initial values with a data-driven initialization scheme. In particular, instead of using only the new observations for the online updating, we run the  algorithm for $\{y_{i}\}_{i=T-k}^{T+H}$, with $k$ small, and choosing a diffuse but proper prior for the initial states at $T-k$. As a result the distribution of the smoothed states at $T$ is not anymore affected by the problem of large conditional variances leading to better online updating performance.

\section{ Simulation Studies}
The aim of the following simulation studies is to compare the performance of our proposed LAF with respect to BCR, and to the models for multivariate stochastic volatility most widely used in practice, specifically: EWMA, PC-GARCH, GO-GARCH and DCC-GARCH. In order to assess whether and to what extent LAF can accommodate, in practice, even sharp changes in the time-varying means and covariances and to evaluate the costs of our flexible approach in settings where the mean and covariance functions do not require locally adaptive estimation techniques, we focus on two different sets of simulated data. The first is based on an underlying structure characterized by locally varying smoothness processes, while the second has means and covariances evolving in time through smooth processes. In the last subsection we also  analyze the performance of the proposed
online updating algorithm.

\subsection{Simulated Data}
\begin{description}
\item{A. Locally varying smoothness processes:} We generate a set of $5$-dimensional observations $y_{i}$ for each $t_{i}$ in the discrete set $\mathcal{T}_{o}=\{1,2,...,100\}$, from the latent factor model in (\ref{eq:2}) with $\Lambda(t_{i})=\Theta\xi(t_{i})$. To allow sharp changes of means and covariances in the generating mechanism, we consider a $2 \times 2$ (i.e. $L=K=2$) matrix $\{\xi(t_{i})\}_{i=1}^{100}$ of time-varying functions adapted from \citet{Dono:1994} with locally-varying smoothness (more specifically we choose `bumps' functions). The latent mean dictionary elements $\{\psi(t_{i})\}_{i=1}^{100}$ are simulated from a Gaussian process $\mbox{GP}(0,c)$ with length scale $\kappa=10$, while the elements in matrix $\Theta$ can be obtained from the shrinkage prior in (\ref{eq:12}) with $a_{1}=a_{2}=10$. Finally the elements of the diagonal matrix $\Sigma_{0}^{-1}$ are sampled independently from $\mbox{Ga}(1,0.1)$.
\item{B. Smooth processes: We consider the same dataset of $10$-dimensional observations $y_{i}$ with $t_{i} \in \mathcal{T}_{o}=\{1,2,...,100\}$ investigated in \citet[section~4.1]{Fox:2011}. The settings are similar to the previous with exception of $\{\xi(t_{i})\}_{i=1}^{100}$ which are $5 \times 4$ (i.e. $L=5, K=4$) matrices of smooth GP dictionary functions with length-scale $\kappa=10$.}
\end{description}

\subsection{Estimation Performance}
\begin{description}
\item{A. Locally varying smoothness processes:}\\
Posterior computation for LAF is performed by using truncation levels $L^{*}=K^{*}=2$ (at higher level settings we found that the shrinkage prior on $\Theta$ results in posterior samples of the elements in the additional columns concentrated around $0$). We place a $\mbox{Ga}(1,0.1)$ prior on the precision parameters $\sigma_{j}^{-2}$ and choose $a_{1}=a_{2}=2$. As regards the nGP prior for each dictionary element $\xi_{lk}(t)$ with $l=1,...,L^{*}$ and $k=1,...,K^{*}$, we choose diffuse but proper priors for the initial values by setting $\sigma^{2}_{\mu_{lk}}=\sigma^{2}_{\alpha_{lk}}=100$ and place an $\mbox{InvGa}(2, 10^{8})$ prior on each $\sigma^{2}_{\xi_{lk}}$ and $\sigma^{2}_{A_{lk}}$ in order to allow less smoothed behavior according to a previous graphical analysis of $\tilde{\Sigma}(t_{i})$ estimated via EWMA. Similarly we set $\sigma^{2}_{\mu_{k}}=\sigma^{2}_{\alpha_{k}}=100$ in the prior for the initial values of the latent state equations resulting from the nGP prior for $\psi_{k}(t)$, and consider $a_{\psi}=a_{B}=b_{\psi}=b_{B}=0.005$ to balance the rough behavior induced on the nonparametric mean functions by the settings of the nGP prior on $\xi_{lk}(t)$, as suggested from previous graphical analysis. Note also that for posterior computation, we first scale the predictor space to $(0,1]$, leading to $\delta_{i}=1/100,$ for $i=1,...,100$.

For inference in BCR we consider the same  previous hyperparameters setting for $\Theta$ and $\Sigma_{0}$ priors as well as the same truncation levels $K^{*}$ and $L^{*}$, while the length scale $\kappa$ in GP prior for $\xi_{lk}(t)$ and $\psi_{k}(t)$ has been set to 10 using the data-driven heuristic outlined in \citet{Fox:2011}. In both cases we run $50{,}000$ Gibbs iterations discarding the first $20{,}000$ as burn-in and thinning the chain every $5$ samples. 

As regards the other approaches, EWMA has been implemented by choosing the smoothing parameter $\lambda$ that minimizes the mean squared error (MSE) between the estimated covariances and the true values. PC-GARCH algorithm follows the steps provided by \citet{Burn:2005} with GARCH(1,1) assumed for the conditional volatilities of each single time series and the principal components. GO-GARCH and DCC-GARCH recall the formulations provided by \citet{Weide:2002} and \citet{Eng:2002} respectively, assuming a GARCH(1,1) for the conditional variances of the processes analyzed, which proves to be a correct choice in many financial applications and also in our setting. Note that, differently from LAF and BCR, the previous approaches do not model explicitly the mean process $\{ \mu(t_{i})\}_{i=1}^{100}$ but work directly on the innovations $\{ y_{i}-\mu(t_{i})\}_{i=1}^{100}$. Therefore in these cases we first model the conditional mean via smoothing spline and in a second step we estimate the models working on the innovations. The smoothing parameter for spline estimation has been set to $0.7$, which was found to be appropriate to best reproduce the true dynamic of $\{ \mu(t_{i})\}_{i=1}^{100}$.
\item{B. Smooth processes:}\\ We mainly keep the same setting of the previous simulation study with few differences. Specifically, $L^{*}$ and $K^{*}$ has been fixed to $5$ and $4$ respectively (also in this case the choice of the truncation levels proves to be appropriate, reproducing the same results provided in the simulation study of \citet{Fox:2011} where $L^{*}=10$ and $K^{*}=10$). Moreover the scale parameters in the Inverse Gamma prior on each $\sigma^{2}_{\xi_{lk}}$ and $\sigma^{2}_{A_{lk}}$ has been set to $10^4$ in order to allow a smoother behavior according to a previous graphical analysis of $\tilde{\Sigma}(t_{i})$ estimated via EWMA, but without forcing the nGP prior to be the same as a GP prior. Following \citet{Fox:2011} we run $10{,}000$ Gibbs iterations which proved to be enough to reach convergence, and discarded the first $5{,}000$ as burn-in.

\end{description}
 In the first set of simulated data,  we analyzed mixing by the Gelman-Rubin procedure (see e.g. \citealp{Gel:1992}), based on potential scale reduction factors computed for each chain by splitting the sampled quantities in $6$ pieces of same length. The analysis shows slower mixing for BCR compared with LAF. Specifically, in LAF  $95\%$ of the chains have a potential reduction factor lower than $1.35$, with a median equal to $1.11$, while in LAF the $95$th quantile is $1.44$ and the median equals $1.18$. Less problematic is the mixing for the second set of simulated data, with potential scale reduction factors having median equal to $1.05$ for both approaches and $95$th quantiles equal to $1.15$ and $1.31$ for LAF and BCR, respectively.

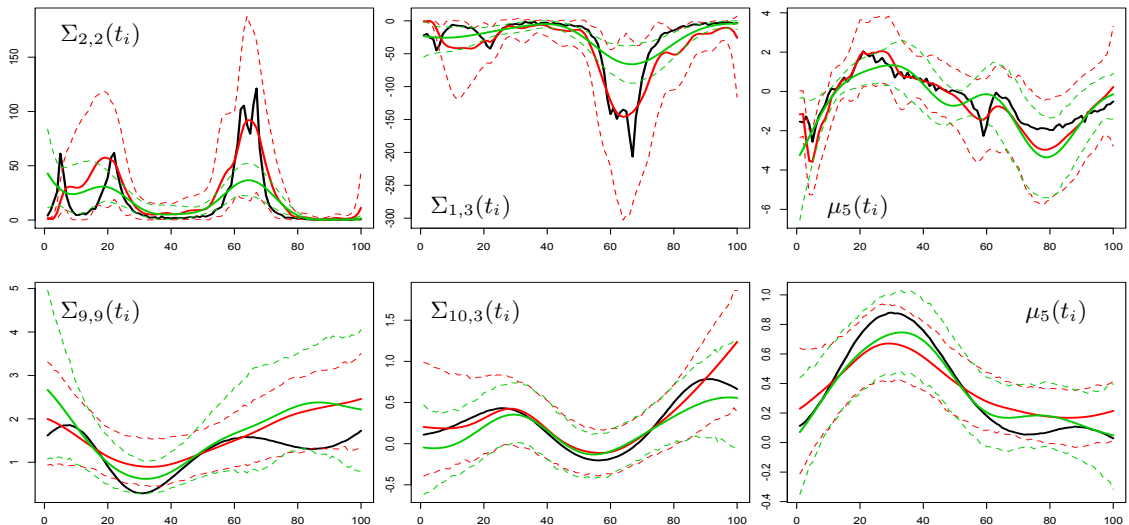
\begin{figure}[t]
\centering
\includegraphics[height=7.3cm, width=15cm]{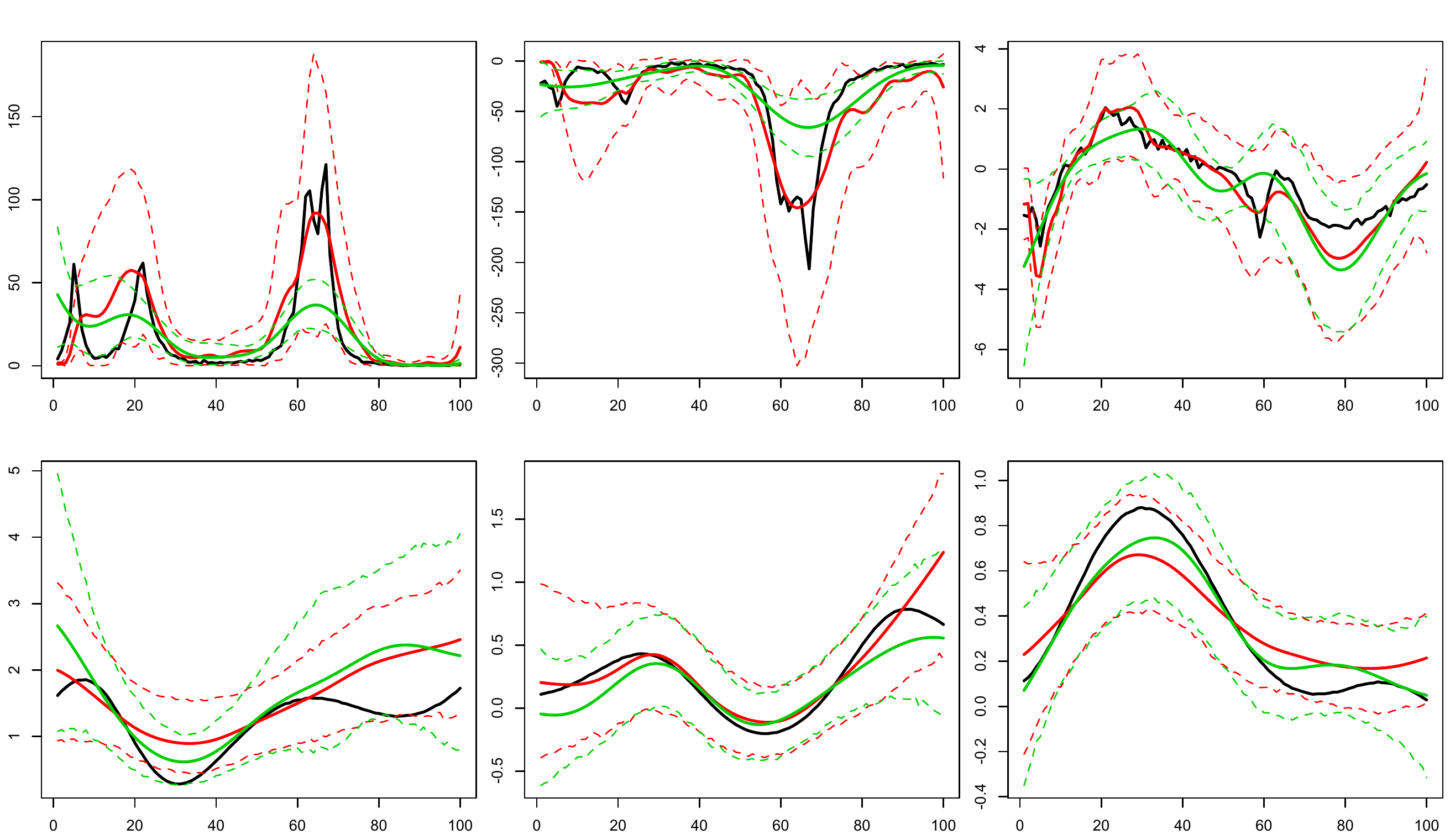} 
\put (-405,185) {{\footnotesize{$\Sigma_{2,2}(t_{i})$}}}
\put (-265,120) {{\footnotesize{$\Sigma_{1,3}(t_{i})$}}}
\put (-115,120) {{\footnotesize{$\mu_{5}(t_{i})$}}}
\put (-405,80) {{\footnotesize{$\Sigma_{9,9}(t_{i})$}}}
\put (-265,80) {{\footnotesize{$\Sigma_{10,3}(t_{i})$}}}
\put (-40,80) {{\footnotesize{$\mu_{5}(t_{i})$}}}
\caption{\footnotesize{For locally varying smoothness simulation (top) and smooth simulation (bottom), plots of truth (black) and posterior mean respectively of LAF (solid red line) and BCR (solid green line) for selected components of the variance (left), covariance (middle), mean (right). For both approaches the dotted lines represent the $95 \%$ highest posterior density intervals.}}
\label{F2}
\end{figure}

Figure~\ref{F2} compares, in both simulated samples, true and posterior mean of the process $\Gamma = \{ \mu(t_i),\Sigma(t_i), i =1,...,100\}$ 
%$\mu(t)$ and $\Sigma(t)
over the predictor space $\mathcal{T}_{o}$ 
together with the point-wise $95\%$ highest posterior density (hpd) intervals for LAF and BCR. From the upper plots we can clearly note that our approach is able to capture conditional heteroscedasticity as well as mean patterns, also in correspondence of sharp changes in the time-varying true functions. The major differences compared to the true values can be found at the beginning and at the end of the series and are likely to be related to the structure of the simulation smoother which also causes a widening of the credibility bands at the very end of the series; for references regarding this issue see \citet{Durb:2001}. However, even in the most problematic cases, the true values are within the bands of the $95\%$ hpd intervals. Much more problematic is the behavior of the posterior distributions for BCR which badly over-smooth both covariance and mean functions leading also to many $95\%$ hpd intervals  not containing the true values. Bottom plots in Figure~\ref{F2} show that the performance of our approach is very close to that of BCR, when data are simulated from a model where the covariances and means evolve smoothly across time and local adaptivity is therefore not required. This happens even if the hyperparameters in LAF are set in order to maintain separation between nGP and GP prior, suggesting large support property for the proposed approach.
\begin{table}[t]
		\centering
		\label{tab:1}

				\begin{tabular}{lcccc}

&Mean &90th Quantile&95th Quantile&Max\\
		\hline
\multicolumn{5}{c}{Covariance $\{\Sigma(t_{i})\}$}\\
\hline
EWMA&$1.37$&$2.28$& $5.49$&$85.86$\\
PC-GARCH&$1.75$&$2.49$&$6.48$&$229.50$\\ 
GO-GARCH&$2.40$&$3.66$&$10.32$&$173.41$\\ 
DCC-GARCH&$1.75$&$2.21$&$6.95$&$226.47$\\ 
BCR&$1.80$&$2.25$&$7.32$&$142.26$\\ 
LAF&$0.90$&$1.99$&$4.52$&$36.95$\\ 
\hline
\multicolumn{5}{c}{Mean $\{\mu(t_{i})\}$}\\
\hline
SPLINE&$0.064$&$0.128$& $0.186$&$2.595$\\
BCR&$0.087$&$0.185$&$0.379$&$2.845$\\ 
LAF&$0.062$&$0.123$&$0.224$&$2.529$\\ 
			\hline
			\end{tabular}
\caption{\footnotesize{LOCALLY VARYING SMOOTHNESS PROCESSES: Summaries of the standardized squared errors between true values $\{\mu(t_{i})\}_{i=1}^{100}$ and $\{\Sigma(t_{i})\}_{i=1}^{100}$ and estimated quantities $\{\hat{\Sigma}(t_{i})\}_{i=1}^{100}$ and $\{\hat{\mu}(t_{i})\}_{i=1}^{100}$ computed with different approaches.}}
\label{tab:1}

	\end{table}
\begin{table}[h!]
		\centering
		
				\begin{tabular}{lcccc}

&Mean &90th Quantile&95th Quantile&Max\\
		\hline
\multicolumn{5}{c}{Covariance $\{\Sigma(t_{i})\}$}\\
\hline
EWMA&$0.030$&$0.081$& $0.133$&$1.119$\\
PC-GARCH&$0.018$&$0.048$&$0.076$&$0.652$\\ 
GO-GARCH&$0.043$&$0.104$&$0.202$&$1.192$\\ 
DCC-GARCH&$0.022$&$0.057$&$0.110$&$0.466$\\ 
BCR&$0.009$&$0.019$&$0.039$&$0.311$\\ 
LAF&$0.009$&$0.022$&$0.044$&$0.474$\\ 
\hline
\multicolumn{5}{c}{Mean $\{\mu(t_{i})\}$}\\
\hline
SPLINE&$0.007$&$0.019$& $0.027$&$0.077$\\
BCR&$0.005$&$0.015$&$0.024$&$0.038$\\ 
LAF&$0.005$&$0.017$&$0.026$&$0.050$\\ 
			\hline
			\end{tabular}
\caption{\footnotesize{SMOOTH PROCESSES: Summaries of the standardized squared errors between true values $\{\mu(t_{i})\}_{i=1}^{100}$ and $\{\Sigma(t_{i})\}_{i=1}^{100}$ and estimated quantities $\{\hat{\Sigma}(t_{i})\}_{i=1}^{100}$ and $\{\hat{\mu}(t_{i})\}_{i=1}^{100}$ computed with different approaches.}}
\label{tab:2}
	\end{table}
\renewcommand{\arraystretch}{1}

The comparison of the summaries of the squared errors between true process $\Gamma = \{ \mu(t_i),\Sigma(t_i), i =1,...,100\}$   and the estimated elements  of % quantities
 $\hat\Gamma=\{\hat{\mu}(t_{i}), \hat{\Sigma}(t_{i}), i=1,\dots,100\}$
%$\{\hat{\mu}(t_{i})\}_{i=1}^{100}$ and $\{\hat{\Sigma}(t_{i})\}_{i=1}^{100}$ 
standardized with the range of the true processes $r_{\mu}=\max_{i,j}\{\mu_{k}(t_{i})\}-\min_{i,j}\{\mu_{j}(t_{i})\}$ and $r_{\Sigma}=\max_{i,j,k}\{\Sigma_{j,k}(t_{i})\}-\min_{i,j,k}\{\Sigma_{j,k}(t_{i})\}$ respectively, once again confirms the overall better performance of our approach relative to all the considered competitors. Table \ref{tab:1} shows that, when local adaptivity is required, LAF provides a superior performance having standardized residuals lower than those of the other approaches.
EWMA seems to provide quite accurate estimates, but it is important to underline that we choose the optimal smoothing parameter $\lambda$ in order to minimize the MSE between estimated and true parameters, which are clearly not known in practical applications. Different values of $\lambda$ reduces significantly the performance of EWMA, which shows also lack of robustness. The closeness of the summaries of LAF and BCR in Table \ref{tab:2} confirms the flexibility of LAF even in settings where local adaptivity is not required and highlights the better performance of the two approaches with respect to the other competitors also when smooth processes are investigated.
\renewcommand{\arraystretch}{.7}

\begin{figure}[t]
\centering
\includegraphics[height=10cm, width=12cm]{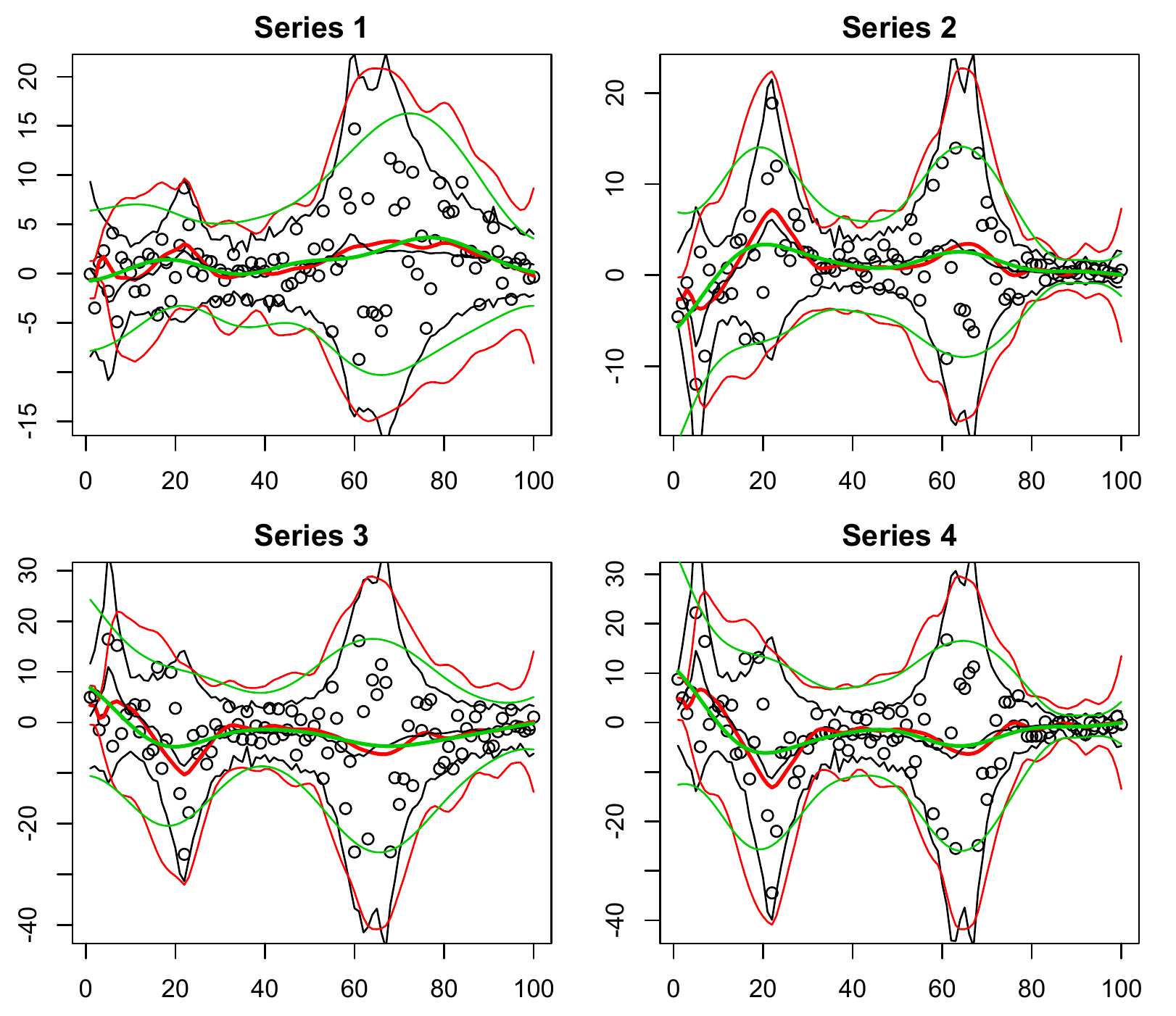}
\caption{\footnotesize{Plot for $4$ selected simulated series of the time-varying mean $\mu_{j}(t_{i})$ and the time-varying $2.5\%$ and $97.5\%$ quantiles of the marginal distribution of $y_{ji}$ with true mean and variance (black), mean and variance from posterior mean of LAF (red), mean and variance from posterior mean of BCR (green). Black points represent the simulated data.}}
\label{F3}
\end{figure}

To better understand the improvement of our approach in allowing locally varying smoothness and to evaluate the consequences of the over-smoothing induced by BCR on the distribution of $y_{i}$ with $i=1,...,100$  consider Figure~\ref{F3} which shows, for some selected series $\{y_{ji}\}_{i=1}^{100}$ in the first simulated dataset, the time varying mean together with the point-wise $2.5\%$ and $97.5\%$ quantiles of the marginal distribution of $y_{ji}$ induced respectively by the true mean and true variance, the posterior mean of $\mu_{j}(t_{i})$ and $\Sigma_{jj}(t_{i})$ from our proposed approach and the posterior mean of the same quantities from BCR. We can clearly see that the marginal distribution of $y_{ji}$ induced by BCR is over-concentrated near the mean, leading to incorrect inferences. Note that our proposal is also able to accommodate heavy tails, a typical characteristic in financial series.

\subsection{Online Updating Performance}
To analyze the performance of the online updating algorithm in LAF model, we simulate $50$ new observations $\{y_{i}\}_{i=101}^{150}$ with $t_{i} \in \mathcal{T}_{o}^{*}=\{101,..., 150\}$, considering the same $\Theta$ and $\Sigma_{0}$ used in the generating mechanism for the first simulated dataset and taking the $50$ subsequent observations of the bumps functions for the dictionary elements $\{\xi(t_{i})\}_{i=101}^{150}$; finally the additional latent mean dictionary elements $\{\psi(t_{i})\}_{i=101}^{150}$ are simulated as before maintaining the continuity with the previously simulated functions $\{\psi(t_{i})\}_{i=1}^{100}$.
According to the algorithm described in subsection $3.3$, we fix $\Theta$, $\Sigma_{0}$, $\Sigma_{\xi}$,  $\Sigma_{A}$,$\Sigma_{\psi}$  and $\Sigma_{B}$  at their posterior mean from the previous Gibbs sampler and consider the last three observations $y_{98}$, $y_{99}$ and $y_{100}$ (i.e. $k=3$) to initialize the simulation smoother in $i=101$ through the proposed data-driven initialization approach. Posterior computation shows  good performance in terms of mixing, and convergence is assessed after $5{,}000$ Gibbs iterations with a  small burn-in of $500$. 

\begin{figure}[t]
\centering
\includegraphics[height=12cm, width=12cm]{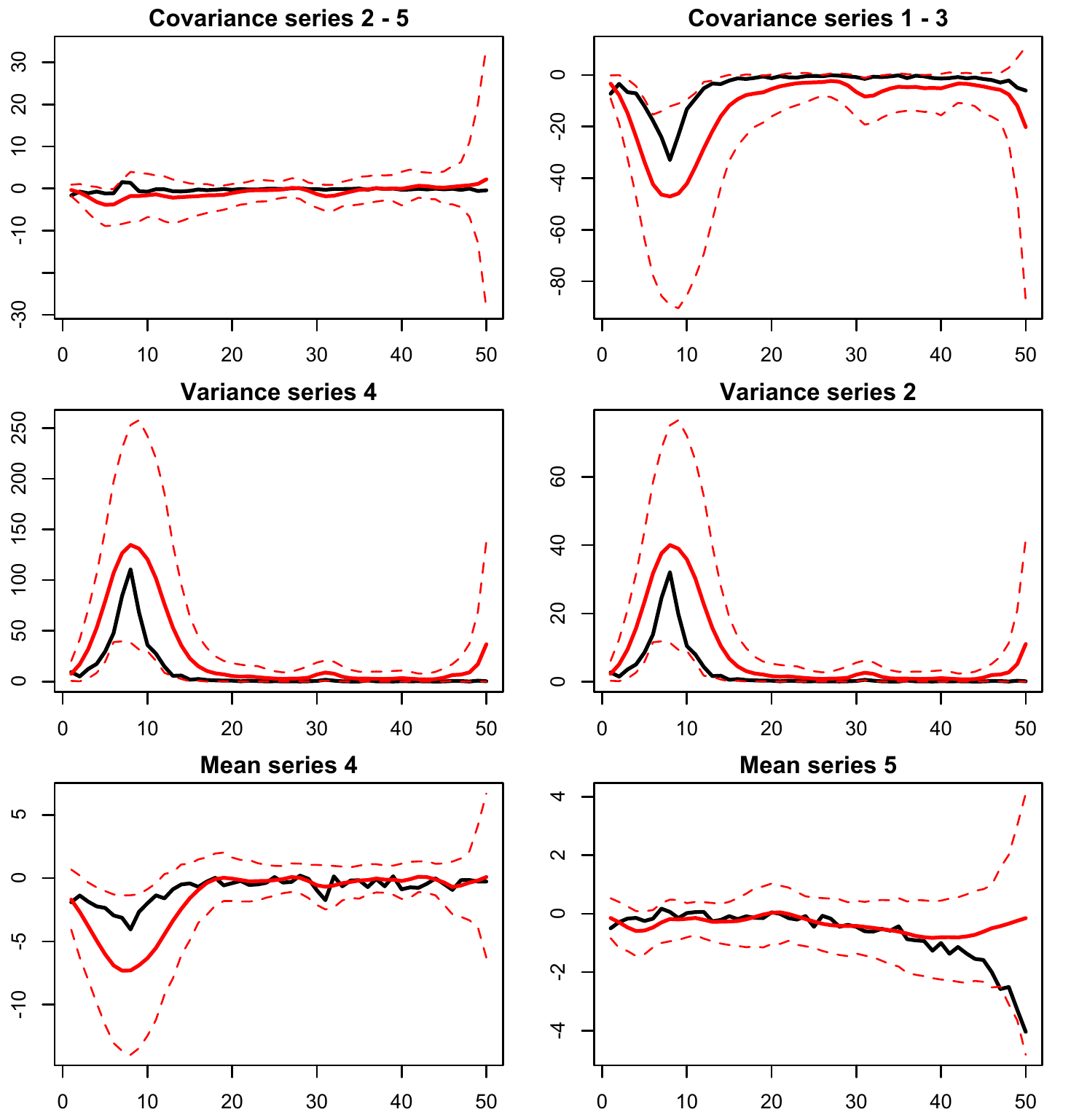}
\caption{\footnotesize{Plots of truth (black) and posterior mean of the online updating procedure (solid red line) for selected components of the covariance (top), variance (middle), mean (bottom). The dotted lines represent the $95 \%$ highest posterior density intervals.}}
\label{F4}
\end{figure}

Figure~\ref{F4} compares true mean and covariance to posterior mean of a select set of components of $\Gamma_*=\{\mu(t_{i}), \Sigma(t_{i}), i=101,...,150\}$
%$\{\mu(t_{i})\}_{i=101}^{150}$ and $\{\Sigma(t_{i})\}_{i=101}^{150}$, 
including also the $95\%$ hpd intervals. The results clearly show that the online updating is characterized by a good performance which allows to capture the behavior of new observations conditioning on the previous estimates. Note that the posterior distribution of the approximated mean and covariance functions tends to slightly over-estimate the patterns of the functions at sharp changes, however also in these cases the true values are within the bands of the credibility intervals. Finally note that the data-driven initialization ensures a good behavior at the beginning of the series, while the results at the end have wider uncertainty bands as expected.

\section{Application Study}
Spurred by the recent growth of interest in the dynamic dependence structure between financial markets in different countries, and in its features during the crises that have followed in recent years, we applied our LAF to the multivariate time series of the main national stock market indices.
\subsection{National Stock Indices (NSI), Introduction and Motivation} 
National Stock Indices represent technical tools that allow, through the synthesis of numerous data on the evolution of the various stocks, to detect underlying trends in the financial market, with reference to a specific basis of currency and time. More specifically, each Market Index can be defined as a weighted sum of the values of a set of national stocks, whose weighting factors is equal to the ratio of its market capitalization in a specific date and overall of the whole set on the same date. 

\begin{figure}[t]
\centering
\includegraphics[height=12cm, width=15cm]{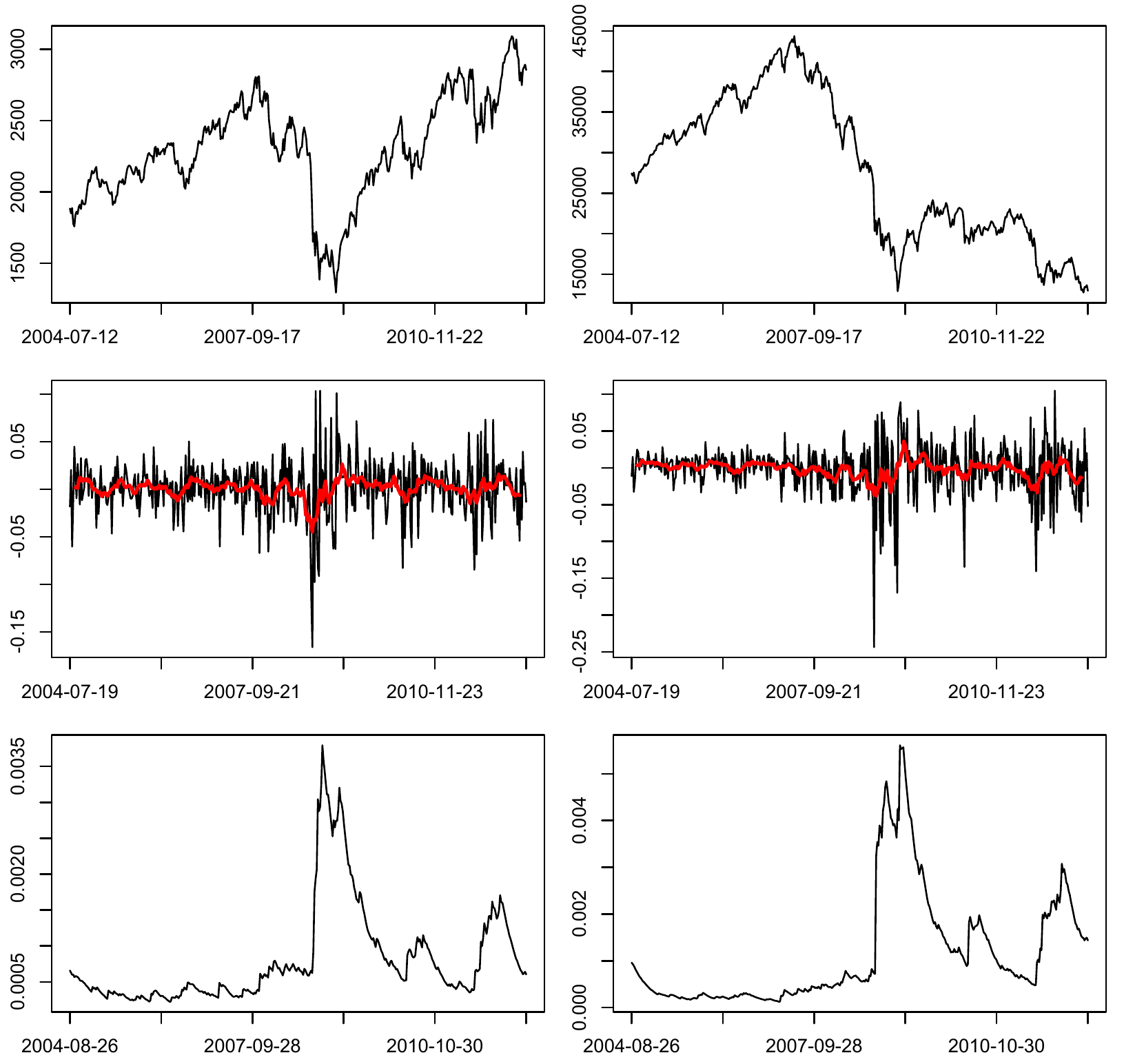}
\put (-405,340) {USA NASDAQ}
\put (-193,340) {ITALY FTSE MIB}
\put (-400,250) {\footnotesize{observed time series}}
\put (-187,250) {\footnotesize{observed time series}}
\put (-400,135) {\footnotesize{log returns}}
\put (-187,135) {\footnotesize{log returns}}
\put (-400,95) {\footnotesize{conditional variances}}
\put (-187,95) {\footnotesize{conditional variances}}
\caption{\footnotesize{Plots of the main features of USA NASDAQ (left) and ITALY FTSE MIB (right). Specifically: observed time series (top), log-returns series (middle) with nonparametric mean estimation via $12$ week Equally Weighted Moving Average (red) in the middle, EWMA volatility estimates (bottom).}}
\label{F5}
\end{figure}
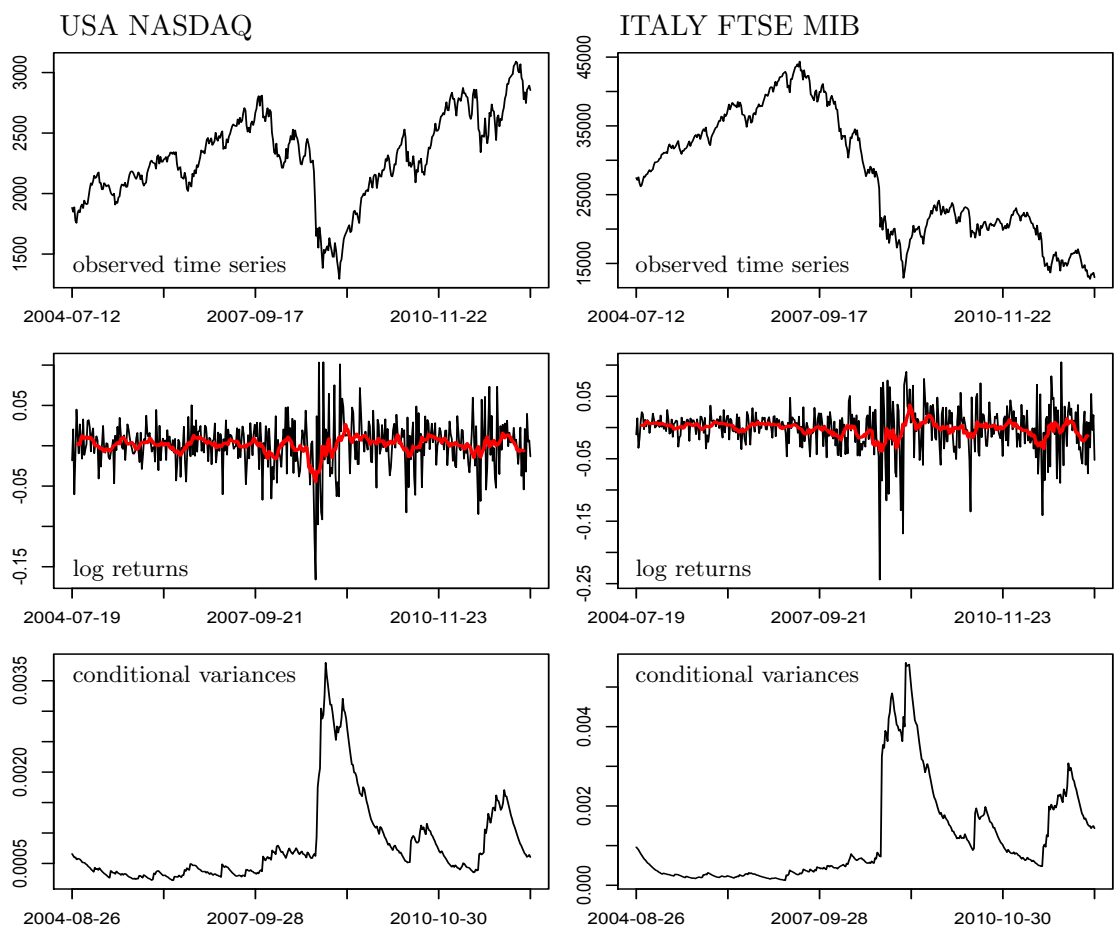

In this application we focus our attention on the multivariate weekly time series of the main $33$ (i.e. $p=33$) National Stock Indices from $12/07/2004$ to $25/06/2012$. Figure~\ref{F5} shows the main features in terms of stationarity, mean patterns and volatility  of two selected NSI downloaded from \url{ http://finance.yahoo.com/}. The non-stationary behavior, together with the different bases of currency and time, motivate the use of logarithmic returns $y_{ji}=\log (I_{ji}/I_{ji-1})$, where $I_{ji}$ is the value of the National Stock Index $j$ at time $t_{i}$. Beside this, the marginal distribution of log returns shows heavy tails and irregular cyclical trends in the nonparametric estimation of the mean, while EWMA estimates highlight rapid changes of volatility during the financial crises observed in the recent years. All these results, together with large settings and high frequency data typical in financial fields, motivate the use of our approach to obtain a better characterization of the time-varying dependence structure among financial markets.

\subsection{LAF for National Stock Index (NSI)} 
We consider the heteroscedastic model $y_{i}\sim \mbox{N}_{33}(\mu(t_{i}),\Sigma(t_{i}))$ for $i=1,...,415$ and $t_{i}$ in the discrete set $\mathcal{T}_{o}=\{1,2,...,415\}$, where the elements of $\Gamma=\{\mu(t_i), \Sigma(t_i), 1=1,...,415\}$, defined by (\ref{subeq3})-(\ref{subeq4}), are induced by the dynamic latent factor model outlined in \ref{eq:2} and \ref{lamb}.

Posterior computation is performed by first rescaling the predictor space $\mathcal{T}_{o}$ to $(0,1]$ and using the same setting of the first simulation study, with the exception of the truncation levels fixed at $K^{*}=4$ and $L^{*}=5$ (which we found to be sufficiently large from the fact that the last few columns of the posterior samples for $\Theta$ assumed values close to $0$) and the hyperparameters of the nGP prior for each $\xi_{lk}(t)$ and $\psi_{k}(t)$ with $l=1,...,L^{*}$ and $k=1,...,K^{*}$, set to $a_{\xi}=a_{A}=a_{\psi}=a_{B}=2$ and $b_{\xi}=b_{A}=b_{\psi}=b_{B}=5 \times 10^7$ to capture also rapid changes in the mean functions according to Figure~\ref{F5}. Missing values in our dataset do not represent a limitation since the Bayesian approach allows us to update our posterior considering solely the observed data. We run $10{,}000$ Gibbs iterations with a burn-in of $2{,}500$. Examination  of trace plots of the posterior samples for
$\Gamma=\{\mu(t_{i}), \Sigma(t_{i}), i=1,...,415\}$ 
%$\{\Sigma(t_{i})\}_{i=1}^{415}$ and $\{\mu(t_{i})\}_{t=1}^{415}$ 
showed no evidence against convergence.

\begin{figure}[t]
\centering
\includegraphics[height=9cm, width=15cm]{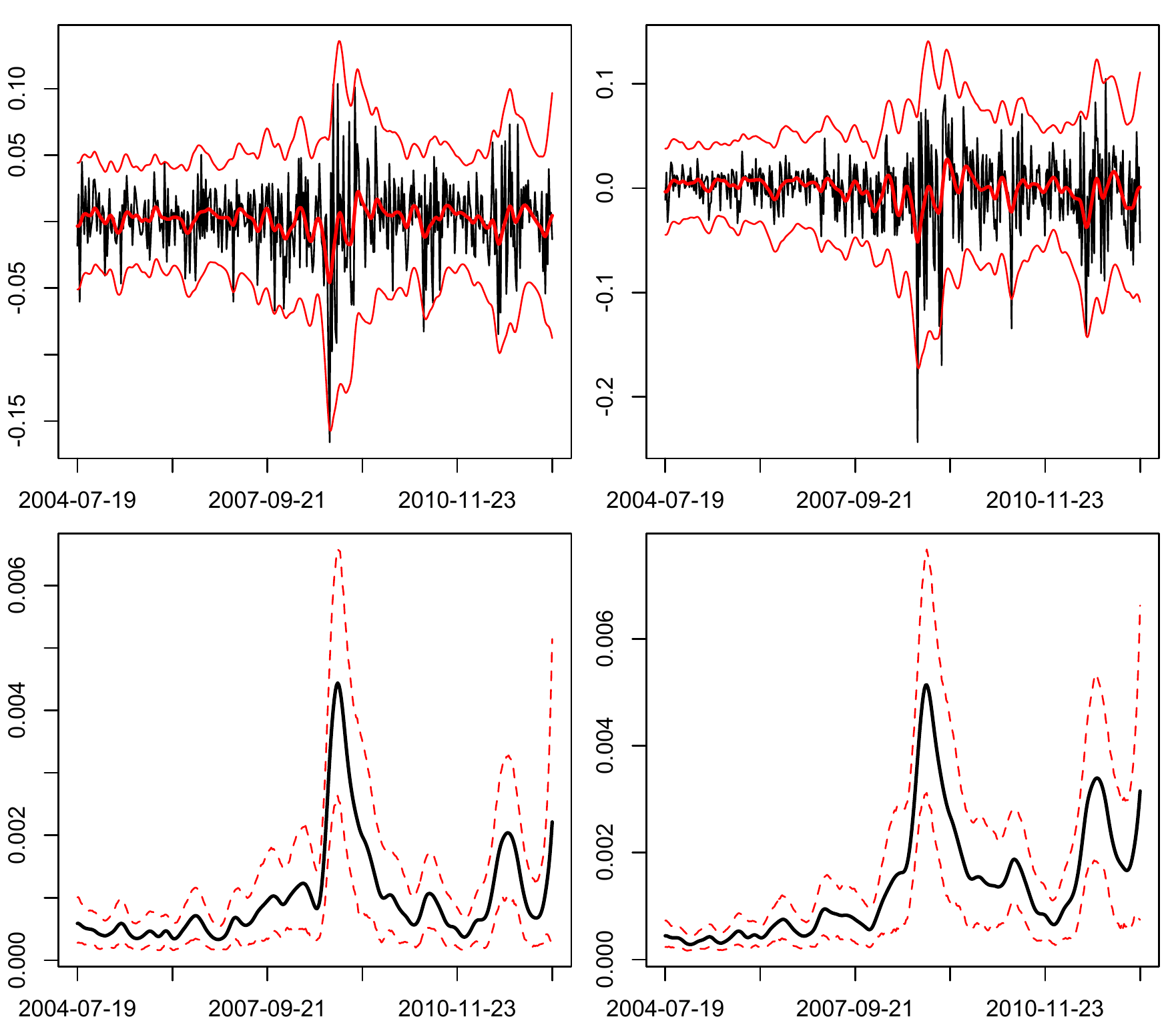}
\put (-405,254) {USA NASDAQ}
\put (-188,254) {ITALY FTSE MIB}
\put (-401,145) {\footnotesize{log returns}}
\put (-185,145) {\footnotesize{log returns}}
\put (-401,110) {\footnotesize{conditional variances}}
\put (-185,110) {\footnotesize{conditional variances}}
\caption{\footnotesize{Top: Plot for $2$ NSI, respectively USA NASDAQ (left) and ITALY FTSE MIB (right), of the log returns (black) and the time-varying estimated mean $\{\hat{\mu}_{j}(t_{i})\}_{i=1}^{415}$ together with the time-varying $2.5\%$ and $97.5\%$ quantiles  (red) of the marginal distribution of $y_{ji}$ from LAF. Bottom: posterior mean (black) and $95\%$ hpd (dotted red) for the variances $\{\Sigma_{jj}(t_{i})\}_{i=1}^{415}$.}}
\label{F6}
\end{figure}

Posterior distributions for the variances in Figure~\ref{F6} demonstrate that we are clearly able to capture the rapid changes in the dynamics of volatility that occur during the world financial crisis of $2008$, in early $2010$ with the Greek debt crisis and in the summer of $2011$ with the financial speculation in government bonds of European countries together with the rejection of the U.S. budget and the downgrading of the United States rating. 
Moreover, the resulting marginal distribution of the log returns induced by the posterior mean of $\mu_{j}(t)$ and $\Sigma_{jj}(t)$, shows that we are also able to accommodate heavy tails as well as mean patterns cycling irregularly between slow and more rapid changes.

\begin{figure}[t]
\centering
\includegraphics[height=10cm, width=13cm]{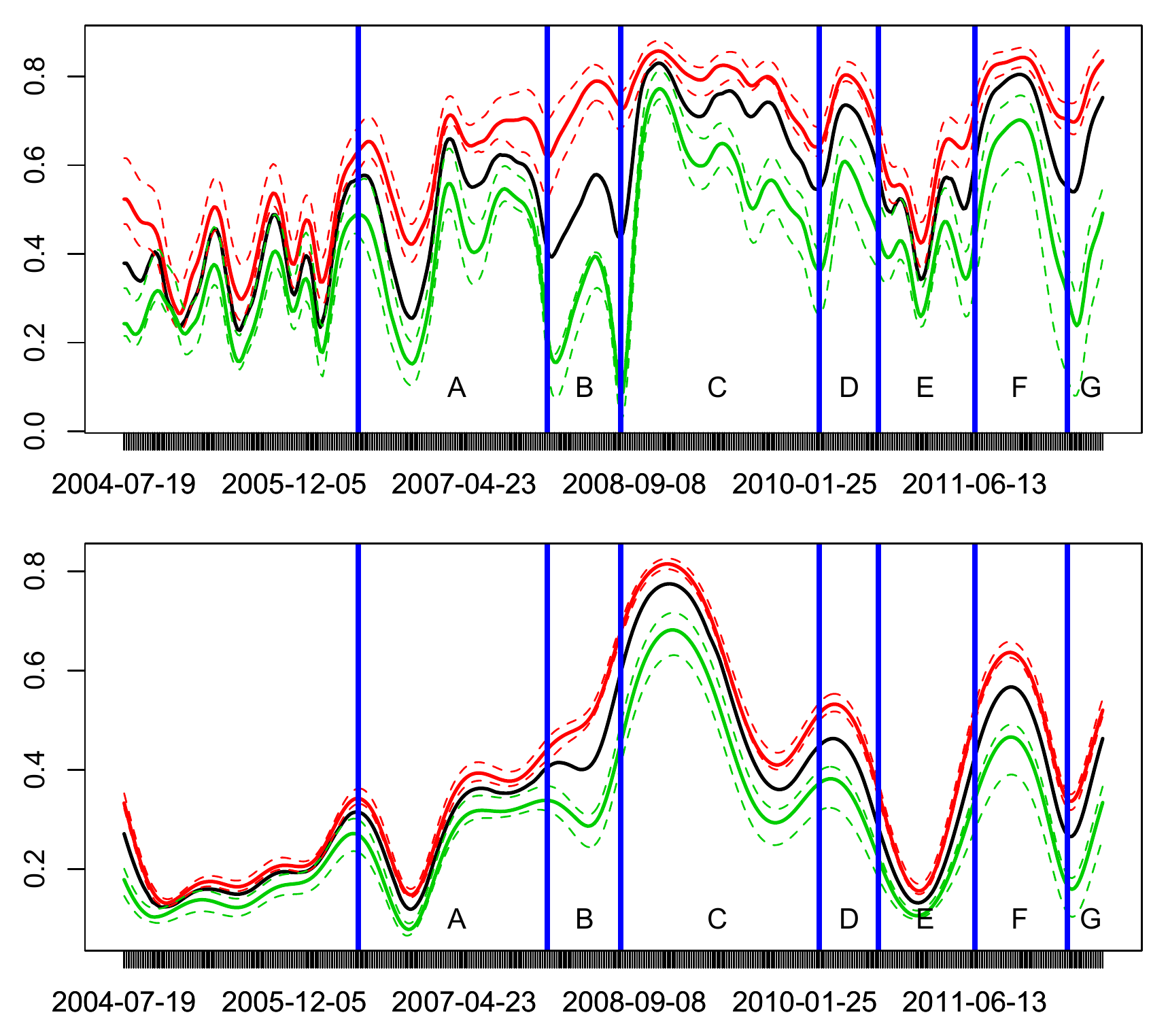}
\put (-320,260) {LAF}
\put (-320,119) {BCR}
\caption{\footnotesize{Black line: For USA NASDAQ median of correlations with the other $32$ NSI based on posterior mean of $\{\Sigma(t_{i})\}_{i=1}^{415}$. Red lines: $25\%$, $75\%$ (dotted lines) and $50\%$ (solid line) quantiles of correlations between USA NASDAQ and European countries (without considering Greece and Russia which present a specific pattern). Green lines: $25\%$, $75\%$ (dotted lines) and $50\%$ (solid line) quantiles of correlations between USA NASDAQ and the countries of Southeast Asia (Asian Tigers and India). Timeline: (A)  burst of U.S. housing bubble; (B) risk of failure of the first U.S. credit agencies (Bear Stearns, Fannie Mae and Freddie Mac); (C) world financial crisis after the Lehman Brothers' bankruptcy; (D) Greek debt crisis; (E) financial reform launched by Barack Obama and EU efforts to save Greece (the two peaks represent respectively Irish debt crisis and Portugal debt crisis); (F) worsening of European sovereign-debt crisis and the rejection of the U.S. budget; (G) crisis of credit institutions in Spain and the growing financial instability of the Eurozone.}}
\label{F7}
\end{figure}

\begin{figure}[t]
\centering
\includegraphics[height=10cm, width=15cm]{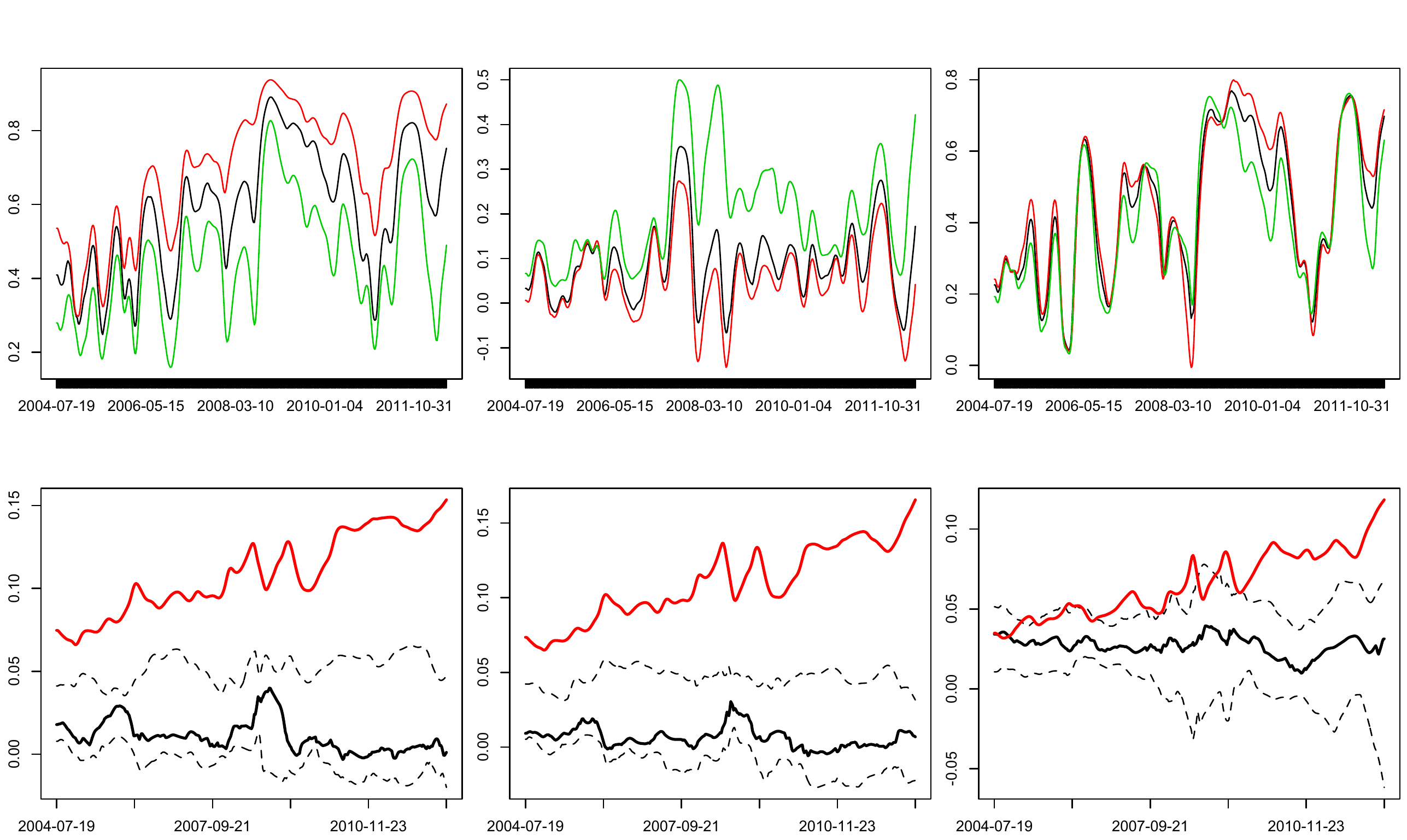}
\put (-410,265) {GERMANY DAX30}
\put (-268,265) {CHINA SSE Composite}
\put (-125,265) {RUSSIA RTSI Index}
\put (-410,123) {ITALY}
\put (-268,123) {SPAIN}
\put (-125,123) {GREECE}
\caption{\footnotesize{Top: For $3$ selected stock market indices, plot of the median of the correlation based on posterior mean of $\{\Sigma(t_{i})\}_{i=1}^{415}$ with the other $32$ world stock indices (black), the European countries without considering Greece and Russia (red) and the Asian Tigers including India (green). Bottom: For $3$ of the European countries more subject to sovereign debt crisis, plot of $25th$, $50th$ and $75th$ quantiles of the time-varying regression parameters based on posterior mean $\{\hat{\Sigma}(t_{i})\}_{i=1}^{415}$ with the other countries (black) and Germany (red).}}
\label{F8}
\end{figure}

Important information about the ability of our model to capture the evolution of world geo-economic structure during different finance scenarios are provided in Figures~\ref{F7} and~\ref{F8}. 
From the correlations between NASDAQ and the other National Stock Indices (based on the posterior mean $\{\hat{\Sigma}(t_{i})\}_{i=1}^{415}$ of the covariances function) in Figure~\ref{F7}, we can immediately notice the presence of a clear geo-economic structure in world financial markets (more evident in LAF than in BCR), where the dependence between the U.S. and European countries is systematically higher than that of South East Asian Nations (Economic Tigers), showing also different reactions to crises. Plots at the top of the Figure~\ref{F8} confirms the above considerations showing how  Western countries exhibit more connection with countries closer in terms of geographical, political and economic structure; the same holds for Eastern countries where we observe a reversal of the colored curves. As expected, Russia is placed in a middle path between the two blocks. A further element that our model captures about the structure of the markets is shown in the plots at the bottom of Figure~\ref{F8}. The time-varying regression coefficients obtained from the standard formulas of the conditional normal distribution based on the posterior mean of  
$\Gamma=\{\mu(t_{i}), \Sigma(t_{i}) , i=1,...,415\}$
%$\{\mu(t_{i})\}_{i=1}^{415}$ and $\{\Sigma(t_{i})\}_{i=1}^{415}$ 
highlight clearly the increasing dependence of European countries with higher crisis in sovereign debt and Germany, which plays a central role in Eurozone as expected.

The flexibility of the proposed approach and the possibility of accommodating varying smoothness in the trajectories over time, allow us to obtain a good characterization of the dynamic dependence structure according with the major theories on financial crisis. Top plot in Figure~\ref{F7} shows how the change of regime in correlations occurs exactly in correspondence to the burst of the U.S. housing bubble (A), in the second half of $2006$. Moreover we can immediately notice that the correlations among financial markets increase significantly during the crises, showing a clear international financial contagion effect in agreement with other theories on financial crisis (see, e.g., \citealp{Bai:1999}, and \citealp{Cl:2009}). As expected the persistence of high levels of correlation is evident during the global financial crisis between late-2008 and end-2009 (C), at the beginning of which our approach also captures a sharp variation in the correlations between the U.S. and Economic Tigers, which lead to levels close to those of Europe. Further rapid changes are identified in correspondence of Greek crisis (D), the worsening of European sovereign-debt crisis and the rejection of the U.S. budget (F) and the recent crisis of credit institutions in Spain together with the growing financial instability Eurozone (G). Finally, even in the period of U.S. financial reform launched by Barack Obama and EU efforts to save Greece (E), we can notice two peaks representing respectively Irish debt crisis and Portugal debt crisis. Note also that BCR, as expected, tends to over-smooth the dynamic dependence structure during the financial crisis, proving to be not able to model the sharp change in the correlations between USA NASDAQ and Economic Tigers during late-2008, and the two peaks representing respectively Irish and Portugal debt crisis at the beginning of 2011.

\subsection{National Stock Indices, Updating and Predicting}
The possibility to quickly update the estimates and the predictions as soon as new data arrive, represents a crucial aspect to obtain quantitative informations about the future scenarios of the crisis in financial markets. To answer this goal, we apply the online updating algorithm presented in subsection $3.3$, to the new set of weekly observations $\{y_{i}\}_{i=416}^{422}$ from $02/07/2012$ to $13/08/2012$ conditioning on posterior estimates of the Gibbs sampler based on observations $\{y_{i}\}_{i=1}^{415}$ available up to $25/06/2012$. We initialized the simulation smoother algorithm with the last $8$ observations of the previous sample. 

\begin{figure}[t]
\centering
\includegraphics[height=10cm, width=15.5cm]{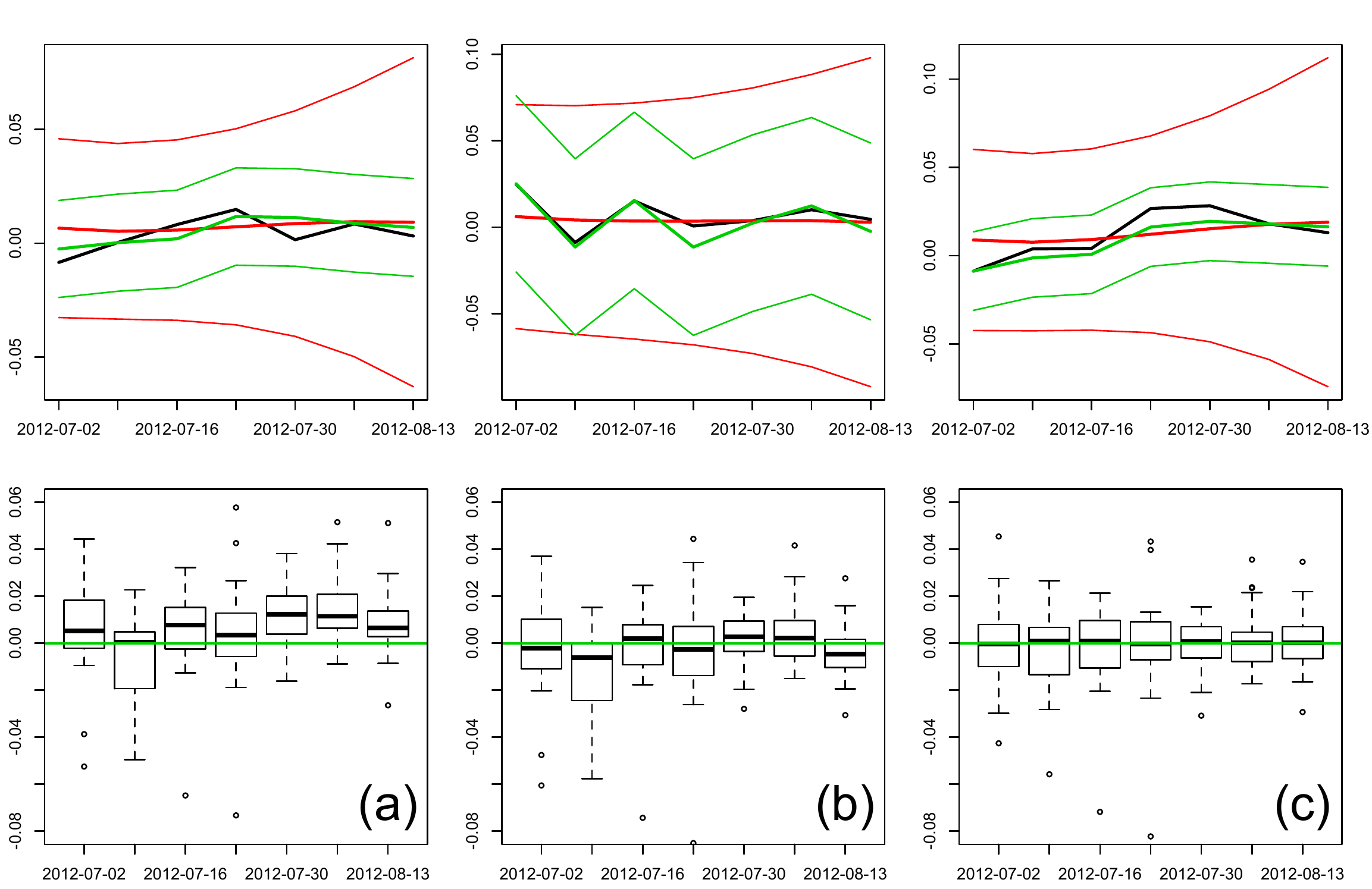}
\put (-422,275) {USA NASDAQ}
\put (-275,275) {INDIA BSE30}
\put (-128,275) {FRANCE CAC40}
\put (-422,130) {\footnotesize{prediction: method (a)}}
\put (-275,130) {\footnotesize{prediction: method (b)}}
\put (-128,130) {\footnotesize{prediction: method (c)}}
\caption{\footnotesize{Top: For $3$ selected NSI, respectively USA NASDAQ (left), INDIA BSE30 (middle) and FRANCE CAC40 (right), plot of the observed log returns (black) together with the mean and the $2.5\%$ and $97.5\%$ quantiles of the marginal distribution (red) and conditional distribution given the other $32$ NSI (green) based on the posterior mean of $\Gamma_*=\{\mu(t_{i}), \Sigma(t_{i}), i=416,...,422\}$ from the online updating procedure for the new observations from $02/07/2012$ to $13/08/2012$. Bottom: boxplots of the one step ahead prediction errors for the $33$ NSI, where the predicted values are respectively: (a) unconditional mean $\{\tilde{y}_{i+1}\}_{i=415}^{421}=0$, (b) marginal mean of the one step ahead predictive distribution using the online updating procedure for $\{\tilde{y}_{i+1|i}\}_{i=415}^{421}$, (c) conditional mean given the log returns of the other $32$ NSI at $i+1$ of the one step ahead predictive distribution using the online updating procedure for $\{\tilde{y}_{i+1|i}\}_{i=415}^{421}$. Predictions for (b) and (c) are induced by the posterior mean of $\{\mu(t_{i+1|i}), \Sigma(t_{i+1|i}), i=415,..,421 \}$ of our LAF.}}
\label{F9}
\end{figure}
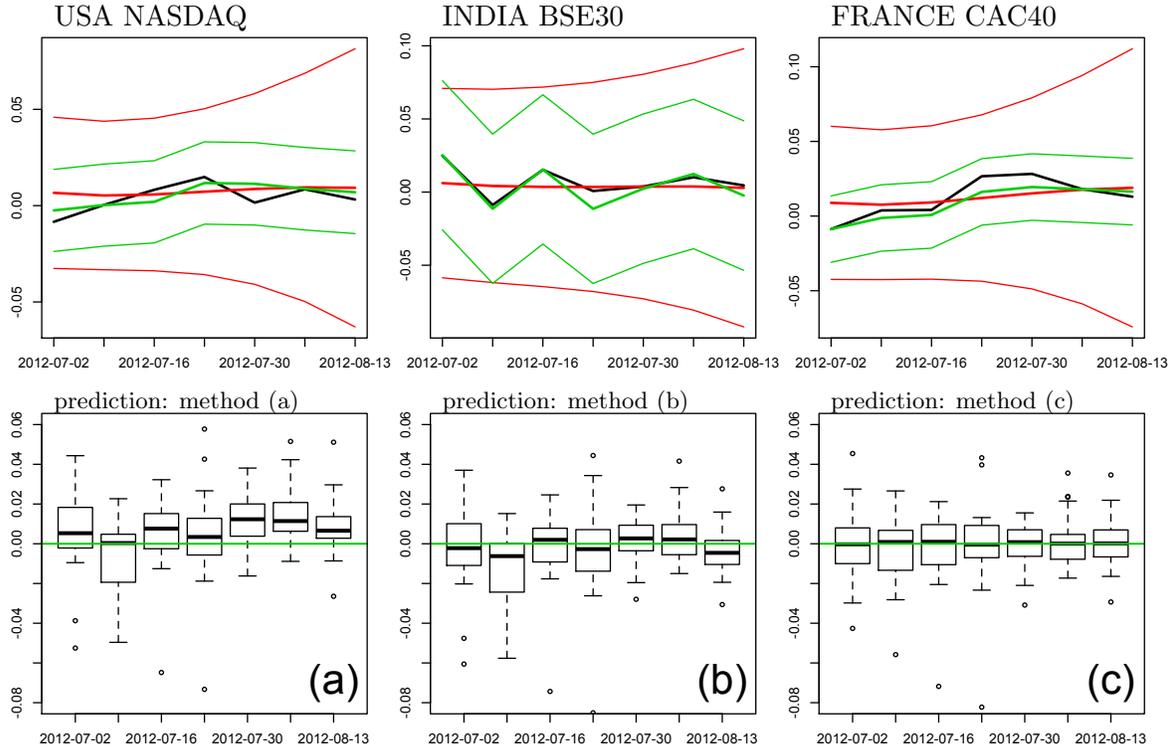

Plots at the top of Figure~\ref{F9} show, for $3$ selected National Stock Indices, the new observed log returns $\{y_{ji}\}_{i=416}^{422}$ (black) together with the mean and the $2.5\%$ and $97.5\%$ quantiles of the marginal distribution (red) and conditional distribution (green) of $y_{ji}|y_{i}^{-j}$ with $y_{i}^{-j}=\{y_{qi},q \neq j\}$.
We use standard formulas of the multivariate normal distribution based on the posterior mean of the updated 
$\Gamma_*=\{\mu(t_{i}), \Sigma(t_{i}), i=416,...,422\}$ 
%$\{\Sigma(t_{i})\}_{i=416}^{422}$ and $\{\mu(t_{i})\}_{i=416}^{422}$ 
after $5{,}000$ Gibbs iterations with a burn in of $500$. This is sufficient for convergence based on examining trace plots of the time-varying mean and covariance matrices. From these results, we can clearly notice the good performance of our proposed online updating algorithm in obtaining a  characterization for the distribution of new observations. Also note that the multivariate approach together with a flexible model for the mean and covariance, allow for significant improvements when the conditional distribution of an index given the others are analyzed.

To obtain further informations about the predictive performance of our LAF, we can easily use our online updating algorithm to obtain $h$ step-ahead predictions for 
$\Gamma_{T+H|T}=\{\mu(t_{T+h|T}), \Sigma(t_{T+h|T}), h=1,...,H\}$.
%$\Sigma(t_{T+h|T})$ and $\mu(t_{T+h|T})$ with $h=1,...,H$. 
In particular, referring to \citet{Durb:2001}, we can generate posterior samples of 
$\Gamma_{T+H|T}$
merely by treating $\{y_{i}\}_{i=T+1}^{T+H}$ as missing values in the proposed online updating algorithm. Here, we consider the one step ahead prediction (i.e. $H=1$) problem for the new observations. More specifically, for each $i$ from $415$ to $421$, we update the mean and covariance functions conditioning on informations up to $t_{i}$ through the online algorithm and then obtain the predicted posterior distribution for 
%$\Gamma_{i+1}=\{\mu(t_{i+1|i}), \Sigma(t_{i+1|i})\}$
$\Sigma(t_{i+1|i})$ and $\mu(t_{i+1|i})$ 
by adding to the sample considered for the online updating a last column $y_{i+1}$ of missing values. 

Plots at the bottom of Figure~\ref{F9}, show the boxplots of the one step ahead prediction errors for the $33$ NSI obtained as the difference between the predicted value $\tilde{y}_{j,i+1|i}$ and, once available, the observed log return $y_{j,i+1}$ with $i+1=416,...,422$ corresponding to weeks from $02/07/2012$ to $13/08/2012$. In (a) we forecast the future log returns with the unconditional mean $\{\tilde{y}_{i+1}\}_{i=415}^{421}=0$, which is what is often done in practice under the general assumption of zero mean, stationary log returns. In (b) we consider $\tilde{y}_{i+1|i}=\hat{\mu}(t_{i+1|i})$, the posterior mean of the one step ahead predictive distribution of $\mu(t_{i+1|i})$, obtained from the previous proposed approach after $5{,}000$ Gibbs iteration with a burn in of $500$. Finally in (c) we suppose that the log returns of all National Stock Indices except that of country $j$ (i.e., $y_{j,i+1}$) become available at $t_{i+1}$ and, considering $y_{i+1|i}\sim N_{p}(\hat{\mu}(t_{i+1|i}),\hat{\Sigma}(t_{i+1|i}))$ with 
%$\hat\Gamma_{i+1}=\{\hat{\mu}(t_{i+1|i}), \hat{\Sigma}(t_{i+1|i})\}$
$\hat{\mu}(t_{i+1|i})$ and $\hat{\Sigma}(t_{i+1|i})$ 
posterior mean of the one step ahead predictive distribution 
respectively for 
%$\Gamma_{i+1}=\{\hat{\mu}(t_{i+1|i}), {\Sigma}(t_{i+1|i})\}$
$\mu(t_{i+1|i})$ and $\Sigma(t_{i+1|i})$, we forecast $\tilde{y}_{j,i+1}$ with the conditional mean of $y_{j,i+1}$ given the other log returns at time $t_{i+1}$. 

Comparing boxplots in (a) with those in (b) we can see that our model allows to obtain improvements also in terms of prediction. Furthermore, by analyzing the boxplots in (c) we can notice how our ability to obtain a good characterization of the time-varying covariance structure can play a crucial role also in improving forecasting, since it enters into the standard formula for calculating the conditional mean in the normal distribution.

\section{ Discussion}
In this paper, we have presented a continuous time multivariate stochastic process for time series to obtain a better characterization for mean and covariance temporal dynamics. Maintaining simple conjugate posterior updates and tractable computations in moderately large $p$ settings, our model increases significantly the flexibility of previous approaches as it captures sharp changes both in mean and covariance dynamics while accommodating heavy tails. Beside these key advantages, the state space formulation enables development of a fast online updating algorithm particularly useful for high frequency data.

The simulation studies highlight the flexibility and the overall better performance of LAF with respect to the models for multivariate stochastic volatility most widely used in practice, both when adaptive estimation techniques are required, and also when the underlying mean and covariance structures do not show sharp changes in their dynamic.

The application to the problem of capturing temporal and geo-economic structure between the main financial markets demonstrates the utility of our approach and the improvements that can be obtained in the analysis of multivariate financial time series with reference to (i) heavy tails, (ii) locally adaptive mean regression, (iii) sharp changes in covariance functions, (iii) high dimensional dataset, (iv) online updating with high frequency data (v) missing values and (vi) predictions.  Potentially further improvements are possible using a stochastic differential equation model that explicitly incorporates prior information on dynamics.

\appendix
\section*{Appendix A1. Posterior Computation}
For a fixed truncation level $L^{*}$ and a latent factor dimension $K^{*}$ the detailed steps of the Gibbs sampler for posterior computations are:
\begin{enumerate}
\item{Define the vector of the latent state and the error terms in the state space equation resulting from nGP prior for dictionary elements as
%{\footnotesize{
\begin{eqnarray*}
\Xi_{i}&=&[\xi_{11}(t_{i}),\xi_{21}(t_{i}),..,\xi_{L^{*}K^{*}}(t_{i}),\xi'_{11}(t_{i})..,\xi'_{L^{*}K^{*}}(t_{i}),A_{11}(t_{i}),..,A_{L^{*}K^{*}}(t_{i})]^T \nonumber \\
\Omega_{i,\xi}&= &[\omega_{i,\xi_{11}},\omega_{i,\xi_{21}},..,\omega_{i,\xi_{L^{*}K^{*}}},\omega_{i,A_{11}},\omega_{i,A_{21}},..,\omega_{i,A_{L^{*}K^{*}}}]^T   
\label{eq:37}
\end{eqnarray*}%}}
Given $\Theta$, $\{\eta_{i}\}_{i=1}^{T}$, $\{y_{i}\}_{i=1}^{T}$, $\Sigma_{0}$ and the variances in latent state equations $\{\sigma^{2}_{\xi_{lk}}\}$,  $\{\sigma^{2}_{A_{lk}}\}$, with $l=1,...,L^{*}$ and $k=1,...,K^{*}$; update $\{\Xi_{i}\}_{i=1}^{T}$ by using the simulation smoother in the following state space model
\begin{eqnarray}
 y_{i}&=&[\eta_{i}^T\otimes\Theta, 0_{p\times (2\times K^{*} \times L^{*})}]\Xi_{i}+\epsilon_{i}
 \label{eq:38}
\\
\Xi_{i+1}&=&T_{i}\Xi_{i}+R_{i}\Omega_{i,\xi}
\label{eq:39}
\end{eqnarray} 
Where the observation equation in (\ref{eq:38}) results by applying the $vec$ operator in the latent factor model $y_{i}=\Theta\xi(t_{i})\eta_{i}+\epsilon_{i}$. More specifically 
recalling the property $vec(ABC)=(C^T\otimes A)vec(B)$ we obtain
\begin{eqnarray*}
y_{i} \ =\ vec(y_{i}) &=& vec \{\Theta\xi(t_{i})\eta_{i}+\epsilon_{i}\}  \nonumber\\
&=& vec\{\Theta\xi(t_{i})\eta_{i}\}+vec(\epsilon_{i}) \nonumber \\
&=& (\eta_{i}^T\otimes\Theta)vec\{\xi(t_{i})\} +\epsilon_{i}.\label{eq:40}
\end{eqnarray*} 
The state equation in (\ref{eq:39}) is a joint representation of the equations resulting from the nGP prior on each $\xi_{lk}$ defined in (\ref{eq:11}). As a result, the $(3 \times L^{*} \times K^{*}) \times (3 \times L^{*} \times K^{*})$ matrix $T_{i}$ together with the $(3 \times L^{*} \times K^{*}) \times (2 \times L^{*} \times K^{*})$ matrix $R_{i}$ reproduce, for each dictionary element the state equation in (\ref{eq:11}) by fixing to $0$ the coefficients relating latent states with different $(l,k)$ (from the independence between the dictionary elements). Finally,  recalling the assumptions on $\omega_{i,\xi_{lk}}$ and $\omega_{i,A_{lk}}$, $\Omega_{i,\xi}$ is normally distributed with $\mbox{E}[\Omega_{i,\xi}]=0$ and $\mbox{E}[\Omega_{i,\xi}\Omega_{i,\xi}^T]=\mbox{diag}(\sigma^{2}_{\xi_{11}}\delta_{i},\sigma^{2}_{\xi_{21}}\delta_{i},...,\sigma^{2}_{\xi_{L^{*}K^{*}}}\delta_{i},\sigma^{2}_{A_{11}}\delta_{i},\sigma^{2}_{A_{21}}\delta_{i},...,\sigma^{2}_{A_{L^{*}K^{*}}}\delta_{i})$.}

\item{Given $\{\Xi_{i}\}_{i=1}^{T}$ sample each $\sigma^{2}_{\xi_{lk}}$ and $\sigma^{2}_{A_{lk}}$ respectively from
\begin{eqnarray*}
\sigma_{\xi_{lk}}^{2}|\{\Xi_{i}\} &\sim& \mbox{InvGa} \left( a_{\xi}+\frac{T}{2},b_{\xi}+\frac{1}{2} \sum^{T-1}_{i=1} \frac{(\xi'_{lk}(t_{i+1})- \xi'_{lk}(t_{i})
-A_{lk}(t_{i})\delta_{i})^{2}}{\delta_{i}} \right) \nonumber\\
\sigma_{A_{lk}}^{2}|\{\Xi_{i}\} &\sim& \mbox{InvGa} \left( a_{A}+\frac{T}{2},b_{A}+\frac{1}{2} \sum^{T-1}_{i=1} \frac{(A_{lk}(t_{i+1})-A_{lk}(t_{i}))^{2}}{\delta_{i}} \right)
\label{eq:41}
\end{eqnarray*} }

\item{Similarly to $\Xi_{i}$ and $\Omega_{i,\xi}$ let
\begin{eqnarray*}
\Psi_{i}&=&[\psi_{1}(t_{i}), \psi_{2}(t_{i}), ...,\psi_{K^{*}}(t_{i}),\psi'_{1}(t_{i}),...,\psi'_{K^{*}}(t_{i}),B_{1}(t_{i}),...,B_{K^{*}}(t_{i})]^T \nonumber \\
\Omega_{i,\psi}&=&[\omega_{i,\psi_{1}},\omega_{i,\psi_{2}},...,\omega_{i,\psi_{K^{*}}},\omega_{i,B_{1}},\omega_{i,B_{2}} ,...,\omega_{i,B_{K^{*}}}]^T \nonumber 
\label{eq:47}
\end{eqnarray*}
be the vectors of the latent state and error terms in the state space equation resulting from nGP prior for $\psi$.
Conditional on $\Theta$, $\{\xi(t_{i})\}_{i=1}^{T}$, $\{y_{i}\}_{i=1}^{T}$, $\Sigma_{0}$, and the variances in latent state equations $\{\sigma^{2}_{\psi_{k}}\}$,  $\{\sigma^{2}_{B_{k}}\}$, with $k=1,...,K^{*}$; sample $\{\Psi_{i}\}_{i=1}^{T}$ from the simulation smoother in the following state space model
\begin{eqnarray}
 y_{i}&=&[\Theta\xi(t_{i}), 0_{p\times (2\times K^{*} )}]\Psi_{i}+\varpi_{i},   \label{eq:48}\\
\Psi_{i+1}&=&G_{i}\Psi_{i}+F_{i}\Omega_{i,\psi},
\label{eq:49}
\end{eqnarray} 
 $\varpi_{i} \sim N(0,\Theta\xi(t_{i})\xi(t_{i})^T\Theta^T+\Sigma_{0})$. The observation equation in (\ref{eq:48}) results by marginalizing out $\nu_{i}$ in the latent factor model with nonparametric mean regression $y_{i}=\Theta\xi(t_{i})\psi(t_{i})+\Theta\xi(t_{i})\nu_{i}+\epsilon_{i}$. Analogously to  $\Xi_{i}$, the state equation in (\ref{eq:49}) is a joint representation of the state equation induced by the nGP prior on each $\psi_{k}$ defined in (\ref{eq:14}); where the $(3 \times K^{*}) \times (3 \times K^{*})$ matrix $G_{i}$ and the $(3 \times K^{*}) \times (2 \times K^{*})$ matrix $F_{i}$ are constructed with the same goal of the matrices $T_{i}$ and $R_{i}$ in the state space model for $\Xi_{i}$. Finally, $\Omega_{i,\psi}\sim N_{2\times K^{*}}(0,\mbox{diag}(\sigma^{2}_{\psi_{1}}\delta_{i},\sigma^{2}_{\psi_{2}}\delta_{i},...,\sigma^{2}_{\psi_{K^{*}}}\delta_{i},\sigma^{2}_{B_{1}}\delta_{i},\sigma^{2}_{B_{2}}\delta_{i},...,\sigma^{2}_{B_{K^{*}}}\delta_{i}))$.}

\item{
Given $\{\Psi_{i}\}_{i=1}^{T}$ update each $\sigma^{2}_{\psi_{k}}$ and $\sigma^{2}_{B_{k}}$ respectively from
%{\footnotesize{
\begin{eqnarray*}
\sigma_{\psi_{k}}^{2}|\{\Psi_{i}\} &\sim& \mbox{InvGa} \left( a_{\psi}+\frac{T}{2},b_{\psi}+\frac{1}{2} \sum^{T-1}_{i=1} \frac{(\psi_{k}'(t_{i+1}) - \psi_{k}'(t_{i})-B_{k}(t_{i})\delta_{i})^{2}}{\delta_{i}} \right) \nonumber \\
\sigma_{B_{k}}^{2}|\{\Psi_{i}\} &\sim& \mbox{InvGa} \left( a_{B}+\frac{T}{2},b_{B}+\frac{1}{2} \sum^{T-1}_{i=1} \frac{(B_{k}(t_{i+1})-B_{k}(t_{i}))^{2}}{\delta_{i}} \right)
\label{eq:50}
\end{eqnarray*}}

\item{
Conditioned on $\Theta$, $\Sigma_{0}$, $y_{i}$, $\xi(t_{i})$ and $\psi(t_{i})$, and recalling $\nu_{i}\sim N_{K^{*}}(0,I_{K^{*}})$; the standard conjugate posterior distribution $\nu_{i}|\Theta, \Sigma_{0}, \tilde{y}_{i}, \xi(t_{i}), \psi(t_{i}) $ is
\begin{eqnarray*}
\mbox{N}_{K^{*}} \left( (I+\xi(t_{i})^T \Theta^T \Sigma_{0}^{-1}\Theta \xi(t_{i}))^{-1} \xi(t_{i})^T\Theta^T\Sigma_{0}^{-1}\tilde{y}_{i}, (I+\xi(t_{i})^T \Theta^T \Sigma_{0}^{-1}\Theta \xi(t_{i}))^{-1}
\right)
\label{eq:51}
\end{eqnarray*}
with $\tilde{y}_{i}=y_{i}-\Theta\xi(t_{i})\psi(t_{i})=\Theta\xi(t_{i})\nu_{i}+\epsilon_{i}$.}

\item{Conditioned on $\Theta$, $\{\eta_{i}\}_{i=1}^{T}$, $\{y_{i}\}_{i=1}^{T}$, and $\{\xi(t_{i})\}_{i=1}^{T}$ (obtained from $\Xi_{i}$), the standard conjugate posterior from which to update $\sigma^{-2}_{j}$ is
\begin{eqnarray*}
\sigma_{j}^{-2}|\Theta, \{\eta_{i}\}, \{y_{i}\}, \{\xi_{t_{i}}\} \sim \mbox{Ga} \left( a_{\sigma}+\frac{T}{2},b_{\sigma}+\frac{1}{2}\sum^{T}_{i=1} (y_{ji}-\theta_{j \cdot}\xi(t_{i})\eta_{i})^{2} \right)
\label{eq:42}
\end{eqnarray*}
Where $\theta_{j \cdot}=[\theta_{j1},..., \theta_{jL^{*}}]$}

\item{Given $\{\eta_{i}\}_{i=1}^{T}$, $\{ y_{i} \}_{i=1}^{T}$, $\{\xi(t_{i})\}_{i=1}^{T}$ and the hyperparameters $\phi$ and $\tau$ the shrinkage prior on $\Theta$ combined with the likelihood for the latent factor model lead to the Gaussian posterior
\renewcommand{\arraystretch}{.5}
\begin{eqnarray*}
\theta_{j\cdot}|\{\eta_{i}\}, \{ y_{i} \}, \{\xi(t_{i}) \}, \phi, \tau \sim \mbox{N}_{L^{*}} \left( \tilde{\Sigma}_{\theta} \tilde{\eta}^T \sigma_{j}^{-2}\left[ \begin{array}{c}
y_{j1}\\
.\\
.\\
.\\
y_{jT}\end{array} \right], \tilde{\Sigma}_{\theta}
\right)
\label{eq:44}
\end{eqnarray*}
\renewcommand{\arraystretch}{1}
where $\tilde{\eta}^T=[\xi(t_{1})\eta_{1},\xi(t_{2})\eta_{2}, ..., \xi(t_{T})\eta_{T}]$ and $$\tilde{\Sigma}^{-1}_{\theta}=\sigma_{j}^{-2}\tilde{\eta}^T\tilde{\eta}+diag(\phi_{j1}\tau_{1},..., \phi_{jL^{*}}\tau_{L^{*}})$$.}
\item{The Gamma prior on the local shrinkage hyperparameter $\phi_{jl}$ implies the standard conjugate posterior given $\theta_{jl}$ and $\tau_{l}$
\begin{eqnarray*}
\phi_{jl}|\theta_{jl},\tau_{l} \sim \mbox{Ga}\left(2, \frac{3+\tau_{l}\theta_{jl}^{2}}{2}
\right)
\label{eq:45}
\end{eqnarray*}}
\item{Conditioned on $\Theta$ and $\tau$, sample the global shrinkage hyperparameters from
\begin{eqnarray*}
\vartheta_{1}|\Theta, \tau^{(-1)} \sim \mbox{Ga} \left(a_{1}+\frac{pL^{*}}{2},1+\frac{1}{2}\sum_{l=1}^{L^{*}}\tau_{l}^{(-1)}\sum_{j=1}^{p}\phi_{jl}\theta_{jl}^{2}\right) \quad \quad \quad \quad \nonumber \\
\vartheta_{h}|\Theta, \tau^{(-h)} \sim \mbox{Ga}\left(a_{2}+\frac{p(L^{*}-h+1)}{2},1+\frac{1}{2}\sum_{l=1}^{L^{*}}\tau_{l}^{(-h)}\sum_{j=1}^{p}\phi_{jl}\theta_{jl}^{2}\right)
\label{eq:46}
\end{eqnarray*}
Where $\tau_{l}^{(-h)}=\prod_{t=1,t\neq h}^{l} \vartheta_{t}$ for $h=1,...,p$}
\item{Given the posterior samples from $\Theta$, $\Sigma_{0}$, $\{ \xi(t_{i})\}_{i=1}^{T}$ and $\{\psi(t_{i})\}_{i=1}^{T}$ the realization of the LAF process for $\{\mu(t_i),\Sigma(t_i), t_i \in \mathcal{T} \}$ conditioned on the data $\{y_{i}\}_{i=1}^{T}$ is 
\begin{eqnarray*}
\mu(t_i)&=&\Theta\xi(t_i)\psi(t_i)\\
\Sigma(t_i)&= &\Theta\xi(t_i)\xi(t_i)^T\Theta^T+\Sigma_{0}.
\end{eqnarray*}}
\end{enumerate}

\section*{Appendix B. Online Updating Algorithm}
Consider $\Theta$, $\Sigma_{0}$, $\{\sigma^{2}_{\xi_{lk}}\}$,  $\{\sigma^{2}_{A_{lk}}\}$, $\{\sigma^{2}_{\psi_{k}}\}$ and  $\{\sigma^{2}_{B_{k}}\}$ fixed at their posterior mean  $\hat{\Theta}$, $\hat{\Sigma}_{0}$, $\{\hat{\sigma}^{2}_{\xi_{lk}}\}$,  $\{\hat{\sigma}^{2}_{A_{lk}}\}$, $\{\hat{\sigma}^{2}_{\psi_{k}}\}$, $\{\hat{\sigma}^{2}_{B_{k}}\}$ respectively, and let $\hat{\Xi}_{T}$, $\hat{\Sigma}_{\Xi_{T}}$ and $\hat{\Psi}_{T}$, $\hat{\Sigma}_{\Psi_{T}}$  be the sample mean and covariance matrix of the posterior distribution respectively for $\Xi_{T}$ and $\Psi_{T}$ obtained from the posterior estimates of the Gibbs sampler conditioned on $\{y_{i}\}_{i=1}^{T}$. 

\begin{enumerate}
\item{Given $\hat{\Theta}$, $\hat{\Sigma}_{0}$, $\{\hat{\sigma}^{2}_{\xi_{lk}}\}$,  $\{\hat{\sigma}^{2}_{A_{lk}}\}$, $\{ \eta_{i} \}_{i=T+1}^{T+H}$ and $\{ y_{i} \}_{i=T+1}^{T+H}$ update $\{\Xi_{i}\}_{i=T+1}^{T+H}$ by using the simulation smoother in the following state space model
\begin{eqnarray*}
 y_{i}&=&[\eta_{i}^T\otimes\hat{\Theta}, 0_{p\times (2\times K^{*} \times L^{*})}]\Xi_{i}+\epsilon_{i} \\
\Xi_{i+1}&=&T_{i}\Xi_{i}+R_{i}\Omega_{i,\xi} 
\label{eq:52}
\end{eqnarray*} 
Where $\Xi_{T+1}$ can be initialized from the standard one step ahead predictive distribution for the state space model $\Xi_{T+1} \sim \mbox{N}(T_{T}\hat{\Xi}_{T},T_{T}\hat{\Sigma}_{\Xi_{T}}T_{T}^T+R_{T}E[\Omega_{T,\xi}\Omega_{T,\xi}^T]R_{T}^T)$}

\item{Conditioned on $\hat{\Theta}$, $\hat{\Sigma}_{0}$, $\{\hat{\sigma}^{2}_{\psi_{k}}\}$,  $\{\hat{\sigma}^{2}_{B_{k}}\}$, $\{ \xi(t_{i}) \}_{i=T+1}^{T+H}$ and $\{ y_{i} \}_{i=T+1}^{T+H}$ sample $\{\Psi_{i}\}_{i=T+1}^{T+H}$ through the simulation smoother in the state space model
\begin{eqnarray*}
 y_{i}&=&[\hat{\Theta}\xi(t_{i}), 0_{p\times (2\times K^{*} )}]\Psi_{i}+\varpi_{i} \\
\Psi_{i+1}&=&G_{i}\Psi_{i}+F_{i}\Omega_{i,\psi} 
\label{eq:53}
\end{eqnarray*} 
Similarly to $\Xi_{T+1}$, $\Psi_{T+1} \sim \mbox{N}(G_{T}\hat{\Psi}_{T},G_{T}\hat{\Sigma}_{\Psi_{T}}G_{T}^T+F_{T}E[\Omega_{T,\psi}\Omega_{T,\psi}^T]F_{T}^T)$}
\\

\item{Given $\hat{\Theta}$, $\hat{\Sigma}_{0}$, $\{y_{i}\}$, $\xi(t_{i})$ and $\psi(t_{i})$, for $i=T+1,...T+H$, sample $\nu_{i}$ from the standard conjugate posterior distribution for $\nu_{i}|\Theta, \Sigma_{0}, \tilde{y}_{i}, \xi(t_{i}), \psi(t_{i})$: 
%{\footnotesize{
\begin{eqnarray*}
\mbox{N}_{K^{*}} \left( (I+\xi(t_{i})^T \Theta^T \Sigma_{0}^{-1}\Theta \xi(t_{i}))^{-1} \xi(t_{i})^T\Theta^T\Sigma_{0}^{-1}\tilde{y}_{i}, (I+\xi(t_{i})^T \Theta^T \Sigma_{0}^{-1}\Theta \xi(t_{i}))^{-1}
\right)
\label{eq:54}
\end{eqnarray*}%}}
with $\tilde{y}_{i}=y_{i}-\Theta\xi(t_{i})\psi(t_{i})=\Theta\xi(t_{i})\nu_{i}+\epsilon_{i}$.}
\\

\item{Compute the updated covariance $\{\Sigma(t_{i})\}_{i=T+1}^{T+H}$ and mean $\{ \mu(t_{i})\}_{i=T+1}^{T+H}$ from the usual equations
\begin{eqnarray*}
\Sigma(t_{i})&=&\hat{\Theta}\xi(t_{i})\xi(t_{i})^T \hat{\Theta}^T+ \hat{\Sigma}_{0}\\
\mu(t_{i})&=&\hat{\Theta}\xi(t_{i})\psi(t_{i}) 
\label{eq:55}
\end{eqnarray*} }
\end{enumerate}

\acks{This research was partially supported by grant R01ES17240 from the National Institute of
Environmental Health Sciences (NIEHS) of the National Institutes of Health (NIH) and by
grant CPDA097208/09 from the University of Padua, Italy. }

\input{bibliografia}

\end{document}

%% file: bibliografia.tex
%\clearpage{\pagestyle{empty}\cleardoublepage}